\documentclass[conference,compsoc]{IEEEtran}

\usepackage{algorithm}
\usepackage{algorithmicx}
\usepackage[noend]{algpseudocode}
\usepackage{array}
\usepackage{amsfonts}
\usepackage{multirow}
\usepackage{enumitem}
\usepackage{makecell}
\usepackage{bm}
\usepackage{amsthm}
\usepackage{setspace}
\usepackage{marvosym}
\usepackage{booktabs}
\usepackage{longtable}
\usepackage{xspace}
\usepackage{tabularray}
\UseTblrLibrary{booktabs}
\setlist{leftmargin=*}
\usepackage{hyperref}
\usepackage[nocompress]{cite}
\usepackage{color}
\usepackage{xcolor}
\usepackage{colortbl}
\usepackage{tabularx}
\usepackage{pifont}

\long\def\comment#1{}

\def\ie{$i.e.$}
\def\eg{$e.g.$}

\newcommand{\partitle}[1]{\vspace{0.3em} \noindent \textbf{#1.}}

\newcommand{\firstpartitle}[1]{\noindent \textbf{#1.}}
\newcommand{\bench}{{\sc LeaFBench}\xspace}
\newcommand{\cmark}{\textcolor{green!60!black}{\checkmark}}  
\newcommand{\xmark}{\textcolor{red}{\ding{55}}}
\newcolumntype{L}[1]{>{\raggedright\arraybackslash}p{#1}}
\hypersetup{
	colorlinks=true,
	linkcolor=blue,
	filecolor=blue,      
	urlcolor=blue,
	citecolor=blue,
}

\newtheorem{definition}{Definition}

\usepackage{tcolorbox}

\newtcolorbox{takeawaybox}{
  colback=gray!20,
  colframe=gray!20,
  coltitle=black,
  arc=4pt,
  boxrule=0.5pt,
  boxsep=2pt,
  left=2pt,
  right=2pt,
  top=2pt,
  bottom=2pt,
  before skip=0.7\baselineskip,
  after skip=0.7\baselineskip
}

\usepackage{tikz}
\newcommand{\chatbox}[2][Templates]{%
\begin{center}
    \begin{tikzpicture}[
            chatbox_inner/.style={
                rectangle, 
                rounded corners, 
                opacity=0, 
                text opacity=1, 
                font=\sffamily\scriptsize,
                text width=0.45\textwidth, 
                text height=9pt, 
                inner xsep=6pt, 
                inner ysep=6pt
            },
           chatbox_prompt_inner/.style={chatbox_inner, align=flush left, xshift=0pt, text height=11pt},
           chatbox_user_inner/.style={chatbox_inner, align=flush left, xshift=0pt},
           chatbox_gpt_inner/.style={chatbox_inner, align=flush left, xshift=0pt},
           chatbox/.style={chatbox_inner, draw=black!25, fill=gray!7, opacity=1, text opacity=0},
           chatbox_prompt/.style={chatbox, align=flush left, fill=gray!1.5, draw=black!30, text height=10pt},
           chatbox_user/.style={chatbox, align=flush left},
           chatbox_gpt/.style={chatbox, align=flush left},
           chatbox2/.style={chatbox_gpt, fill=green!25},
           chatbox3/.style={chatbox_gpt, fill=red!20, draw=black!20},
           chatbox4/.style={chatbox_gpt, fill=yellow!30},
           labelbox/.style={
           rectangle, 
           rounded corners, 
           draw=black!50, 
           font=\sffamily\scriptsize\bfseries, 
           fill=gray!5, 
           inner sep=3pt
           },
        ]

        \node[chatbox_user] (q1)[align=justify, text width=0.45\textwidth] {#2};
        \node[chatbox_user_inner] (q1_text)[align=justify, text width=0.45\textwidth] at (q1) {#2};
        \node[labelbox, anchor=north west, yshift=5pt, xshift=5pt] at (q1.north west) {\textbf{#1}};
    \end{tikzpicture}
\end{center}
}

\definecolor{level1}{RGB}{255, 249, 196}  
\definecolor{level2}{RGB}{255, 241, 118}  
\definecolor{level3}{RGB}{220, 231, 117}  
\definecolor{level4}{RGB}{174, 213, 129}  
\definecolor{level5}{RGB}{139, 195, 74}   

\newcommand{\colorauc}[1]{%
  \ifdim #1pt < 0.6pt \cellcolor{level1!80}#1\else
  \ifdim #1pt < 0.7pt \cellcolor{level2!70}#1\else
  \ifdim #1pt < 0.8pt \cellcolor{level3!70}#1\else
  \ifdim #1pt < 0.9pt \cellcolor{level4!70}#1\else
  \cellcolor{level5!70}#1\fi\fi\fi\fi
}

\newcommand{\colormd}[1]{%
  \ifdim #1pt < 0.5pt \cellcolor{level1!80}#1\else
  \ifdim #1pt < 1.0pt \cellcolor{level2!80}#1\else
  \ifdim #1pt < 2.0pt \cellcolor{level3!80}#1\else
  \ifdim #1pt < 2.5pt \cellcolor{level4!80}#1\else
  \cellcolor{level5!80}#1\fi\fi\fi\fi
}

\usepackage[cmex10]{amsmath}
\usepackage{caption}
\usepackage[font=footnotesize]{subfig}

\begin{document}

\date{}

\title{SoK: Large Language Model Copyright Auditing via Fingerprinting}

\author{
{\rm Shuo Shao\textsuperscript{1}, Yiming Li\textsuperscript{3,\Letter}, Yu He\textsuperscript{1}, Hongwei Yao\textsuperscript{4}, Wenyuan Yang\textsuperscript{5}, Dacheng Tao\textsuperscript{3}, Zhan Qin\textsuperscript{1,2}} \\
\textsuperscript{1}The State Key Laboratory of Blockchain and Data Security, Zhejiang University \\
\textsuperscript{2}Hangzhou High-Tech Zone (Binjiang) Institute of Blockchain and Data Security \\ 
\textsuperscript{3}Nanyang Technological University,
\textsuperscript{4}City University of Hong Kong,
\textsuperscript{5}Sun Yat-Sen University\\
{\tt \small\{shaoshuo\_ss, yuherin, qinzhan\}@zju.edu.cn; liyiming.tech@gmail.com; }\\ 
{\tt \small yao.hongwei@cityu.edu.hk; yangwy56@mail.sysu.edu.cn; dacheng.tao@ntu.edu.sg}
} 

\maketitle

\begin{abstract}

The broad capabilities and substantial resources required to train Large Language Models (LLMs) make them valuable intellectual property, yet they remain vulnerable to copyright infringement, such as unauthorized use and model theft. LLM fingerprinting, a non-intrusive technique that compares the distinctive features (\ie, fingerprint) of LLMs to identify whether an LLM is derived from another, offers a promising solution to copyright auditing. However, its reliability remains uncertain due to the prevalence of diverse model modifications and the lack of standardized evaluation. In this SoK, we present the first comprehensive study of the emerging LLM fingerprinting. We introduce a unified framework and taxonomy that structures the field: white-box methods are classified based on their feature source as static, forward-pass, or backward-pass fingerprinting, while black-box methods are distinguished by their query strategy as either untargeted or targeted. Furthermore, we propose \bench, the first systematic benchmark for evaluating LLM fingerprinting under realistic deployment scenarios. Built upon 7 mainstream foundation models and comprising 149 distinct model instances, \bench integrates 13 representative post-development techniques, spanning both parameter-altering methods (\eg, fine-tuning, quantization) and parameter-independent techniques (\eg, system prompts, RAG). Extensive experiments on \bench reveal the strengths and weaknesses of existing methods, thereby outlining future research directions and critical open problems in this emerging field. The code is available at \url{https://github.com/shaoshuo-ss/LeaFBench}.
\end{abstract}

\section{Introduction}
\label{sec:introduction}


With the rapid advancement of Large Language Models (LLMs), they have been extensively deployed across diverse domains, including content creation~\cite{qwen2.5}, language translation~\cite{grattafiori2024llama3}, and code generation~\cite{he2025benchmarking}. Compared with traditional small models, the development of LLMs is considerably more resource-intensive, demanding large-scale datasets and computational resources~\cite{li2025rethinking, qwen2.5}. However, despite their high value, LLMs are increasingly exposed to serious copyright infringement risks. For open-source models, adversaries may exploit them in unauthorized contexts (\eg, prohibited commercial applications) or fine-tune the models and falsely claim independent authorship~\cite{yoon2025intrinsic}. For closed-source models, adversaries can also extract model parameters or replicate functionality through model stealing attacks~\cite{carlini2024stealing, liu2025model}. These threats highlight an urgent need for effective mechanisms to protect and audit the copyright of LLMs in practice.


LLM copyright auditing aims to determine whether a \emph{suspicious model} is derived from a copyright-reserved \emph{source model}. Existing techniques can be broadly categorized into two classes: LLM watermarking and fingerprinting~\cite{sun2023deep}. LLM watermarking\footnote{Some existing works embed ``fingerprints'' into LLMs by modifying their parameters~\cite{xu2024instructional, yamabe2025mergeprint, cai2025utf}. In this paper, consistent with the classic narrow definitions of model watermarking and fingerprinting~\cite{xue2021intellectual, shao2025explanation, li2025move}, we classify such methods as LLM watermarking.} refers to the intrusive injection of a specific identifier (\ie, watermark) into the model~\cite{xu2024instructional, krauss2024clearstamp}, which can later be extracted and validated to verify ownership. While effective in principle, watermarking requires direct modification of the model (\eg, fine-tuning), which may not only degrade model performance but also incur substantial computational overhead. More importantly, watermarking cannot be applied to protect LLMs that have already been released, substantially limiting its applicability in real-world auditing scenarios.


To overcome the limitations of watermarking, LLM fingerprinting techniques have recently emerged~\cite{pasquini2025llmmap, zhang2025reef, zeng2024huref, gubri2024trap}. Fingerprinting involves non-intrusively extracting distinctive fingerprints (\eg, parameter statistics or output characteristics) from the models. By comparing the fingerprints, one can assess the similarity of different LLMs, and a high degree of similarity often indicates copyright infringement. Unlike watermarking, fingerprinting does not require modifying models. As such, it has no impact on the model performance and is also feasible for auditing released models. Notably, LLM fingerprinting has already seen practical application. For example, it was employed to reveal that a student group at Stanford had plagiarized the MiniCPM model~\cite{hu2024minicpm}. More recently, several studies suggested that a famous open-source MoE model may have been developed based on Qwen-2.5~\cite{yoon2025intrinsic, zhang2025matrix}, attracting widespread attention. These real-world cases highlight the practical potential of LLM fingerprinting for copyright protection and auditing.


Despite its promise, designing an effective and reliable LLM fingerprinting method is non-trivial. In contrast to fingerprinting classification models~\cite{chen2022teacher, chen2022copy, shao2025fit}, LLM fingerprinting faces two fundamental challenges: \textbf{(1)} The generation processes of LLMs are inherently non-deterministic, and such randomness may hinder the extraction of stable and consistent fingerprints. \textbf{(2)} Modern LLMs typically incorporate various techniques and components before deployment, including parameter-altering approaches (\eg, fine-tuning and quantization~\cite{jin2024comprehensive}) as well as parameter-independent mechanisms (\eg, system prompts~\cite{giray2023prompt}, and retrieval-augmented generation (RAG)~\cite{lewis2020retrieval}). These factors significantly complicate the reliability of fingerprinting. Consequently, while fingerprinting has demonstrated potential, a critical question remains:  \emph{Are existing LLM fingerprinting methods truly reliable in practice?}

In this SoK, we present the first comprehensive study of the emerging LLM copyright auditing via fingerprinting, aiming to provide a unified framework for analyzing existing and future methods. We formalize LLM fingerprinting as a two-stage framework: an $\rm{Extract}$ function that extracts a fingerprint from a source model, and a $\rm{Verify}$ function that determines whether the fingerprint is present in a suspicious model. This allows us to construct a clear taxonomy that systematizes the current landscape. Specifically, we categorize methods by the auditor's access into \emph{white-box} and \emph{black-box} approaches. White-box methods are further classified into \emph{static}, \emph{forward-pass}, and \emph{backward-pass} techniques, according to their feature source. Black-box methods, in contrast, are divided into \emph{untargeted} and \emph{targeted} fingerprinting, depending on whether they rely on specific query–response pairs. Leveraging this taxonomy, we systematically review existing LLM fingerprinting methods, clarifying the connections and distinctions between different techniques.

Beyond reviewing existing works, we address the lack of standardized evaluation by introducing the \underline{L}arg\underline{e} L\underline{a}nguage Model \underline{F}ingerprinting \underline{Bench}mark (\bench), the first systematic benchmark for evaluating LLM fingerprinting. \bench comprises 149 carefully constructed LLM instances, derived from various mainstream base models. It further integrates 13 representative post-development techniques, ranging from parameter-altering methods (\eg, fine-tuning and model merging) to parameter-independent optimizations (\eg, RAG and CoT). This design enables fair and comprehensive evaluation of fingerprinting methods under realistic conditions. Using \bench, we conduct a comprehensive experimental study of 8 state-of-the-art (SOTA) fingerprinting techniques. Our empirical analysis reveals important insights into the strengths and limitations of existing methods, highlights promising research directions, and uncovers critical open problems for advancing LLM copyright auditing in this rapidly evolving domain.

The contributions of this paper are three-fold.

\begin{itemize}
    \item \textbf{Unified Framework and Taxonomy}. We establish a principled foundation for LLM fingerprinting by proposing a formal framework, defining key objectives, and constructing a clear taxonomy. This structure enables a systematic review and categorization of existing techniques.
    \item \textbf{Benchmark for Standardized Evaluation}. We develop \bench, the first systematic benchmark dedicated to LLM fingerprinting. It provides a standardized toolbox to evaluate fingerprinting methods under a diverse range of representative and realistic model modifications.
    \item \textbf{Comprehensive Empirical Study}. We conduct extensive experiments on 8 existing representative and SOTA LLM fingerprinting methods using our \bench. Our analysis summarizes key insights into the effectiveness and limitations of existing solutions, explores their practical applicability, and identifies critical open challenges and future research directions.
\end{itemize}


\section{Landscape of LLM Copyright Auditing}

\subsection{Background}

\firstpartitle{Large Language Models (LLM)} An LLM is a complex computational system, typically built on the Transformer architecture and characterized by its massive parameter scale and auto-regressive generation paradigm~\cite{grattafiori2024llama3}. Its development generally follows a multi-stage pipeline. It begins with pre-training, during which the model is trained on large-scale corpora to acquire general linguistic knowledge and syntactic patterns. After pre-training, developers apply additional techniques to adapt or refine the model for specific tasks or deployment environments. Such adaptations may involve directly modifying the model parameters through approaches like fine-tuning~\cite{hu2022lora} and compression~\cite{jin2024comprehensive}. Alternatively, they may steer model behavior at inference time using parameter-independent mechanisms, such as utilizing different system prompts~\cite{giray2023prompt} or a retrieval-augmented generation (RAG) mechanism~\cite{lewis2020retrieval}.

\partitle{LLM Copyright Infringements} Developing LLMs requires large-scale datasets and massive computational resources, leading LLMs to valuable digital assets for their owner~\cite{li2025rethinking}. However, these valuable LLMs face significant copyright infringement risks in their application. An adversary could illegally utilize an open-source model in unauthorized areas or fine-tune it and claim the fine-tuned version as their self-developed model~\cite{yoon2025intrinsic}. For instance, a student group is reported to have plagiarized the MiniCPM model~\cite{hu2024minicpm}. For closed-source models that do not directly disclose their model parameters, this threat also has not been alleviated. Existing study~\cite{carlini2024stealing} also demonstrates that it is possible for an adversary to extract precise information from production black-box LLMs. These real-world examples of illicit reuse and direct parameter theft underscore that such threats are not merely theoretical. Therefore, protecting and auditing the copyright of LLMs has become critically important in LLM applications~\cite{ren2024sok}.

\subsection{Problem Formulation}

In this section, we formalize the problem of LLM copyright auditing. Consider a model owner (or auditor) who develops an LLM and subsequently makes it available either by releasing it on an open-source platform such as Hugging Face or by deploying it as a cloud service. An adversary, however, may obtain the model's parameters through direct copying or model stealing. The adversary can then redeploy the stolen model in their own applications while falsely and maliciously claiming independent authorship.


The primary objective of an LLM copyright auditing method is to determine whether a suspicious third-party model is a direct copy of, or a derivative work based on, the developer's source model. Formally, the problem of LLM copyright auditing can be defined as follows.

\begin{definition}[LLM Copyright Auditing]
\label{def:audit}
    Given a suspicious model $M_s$ and the source model $M_o$, an ideal LLM copyright auditing method $\mathcal{A}$ outputs 1 if $M_s$ is indeed developed from $M_o$, otherwise 0, as follows.
    \begin{equation}
    \label{eq:audit}
        \mathcal{A}(M_s, M_o)=\Big\{
        \begin{aligned}
        1,&\enspace M_s\text{ is derived from }M_o \\
        0,&\text{ otherwise}
        \end{aligned}.
    \end{equation}
\end{definition}

In this context, a model $M_s$ is considered ``derived from'' $M_o$ if it inherits a substantial portion of its learned parameters, and consequently its functional behavior, from $M_o$. This relationship holds regardless of whether $M_s$ is a direct copy, an illicitly stolen model, or a subsequent model created by applying post-development modifications (\eg, fine-tuning, quantization) to $M_o$. Collectively, the base model and its derivatives form a ``model lineage''. The primary goal of LLM copyright auditing is to distinguish the independently developed models from true derivatives.

Notably, the method for identifying such ``suspicious models'' (and then auditing them) is generally not included in LLM copyright auditing~\cite{pasquini2025llmmap, zhang2024remark}. The identification may rely on non-technical indicators, such as observing similarities in performance, functionality, or parameter count. This area is beyond the scope of our SoK.

\subsection{Threat Model}

There are two parties involved in the threat model of LLM copyright auditing, including the adversary and the model owner (or auditor).


\partitle{Assumptions of the Adversary} The adversary seeks to improperly appropriate a model (referred to as the \emph{source model}) originally developed by another model owner by copying and deploying it for unauthorized purposes or falsely claiming independent authorship. The model released or deployed by the potential adversary is referred to as the \emph{suspicious model}. Once in possession of the source model after pre-training, the adversary has full control over it and could freely apply different techniques to adapt or optimize the model. The techniques will be discussed in Section~\ref{sec:technique}.


\partitle{Assumptions of the Model Owner/Auditor} The fundamental goal for the model owner or an auditor is to detect whether a suspicious model is an illicit copy or derivative of their original source model. To this end, the owner/auditor is assumed to have the following capabilities:

\begin{itemize}[leftmargin=*]
    \item \textbf{Full control over the source model.} The owner/auditor inherently possesses complete access to their own model $M_o$. This includes its architecture, parameters, and potentially the original training data. They can interact with and modify the model without any restrictions.
    \item \textbf{Access to the suspicious model.} The owner/auditor has access to the suspicious model. This access may be white-box, allowing inspection of the suspicious model's internal parameters, or black-box, where interaction is limited to API queries and observing the corresponding outputs.
\end{itemize}

\subsection{LLM Copyright Auditing Paradigms}

To fulfill LLM copyright auditing, two primary technical paradigms have emerged in the literature: LLM watermarking~\cite{shao2025explanation, xu2024instructional, xu2025mark} and LLM fingerprinting~\cite{pasquini2025llmmap, yang2024fingerprint, mcgovern2025your}. These approaches differ fundamentally in whether they require modification of the model being protected.

\partitle{LLM Watermarking} LLM watermarking~\cite{shao2025explanation, xu2024instructional, xu2025mark} is an active approach that intrusively embeds a unique identifier (\ie, watermark) into the source model $M_o$ before its release or deployment. This process involves a pair of algorithms: an embedding function and a verification function. The copyright can later be verified by checking for the presence of this embedded watermark in a suspicious model $M_s$. The formal definition is as follows.

\begin{definition}[LLM Watermarking]
LLM watermarking involves embedding a specific signal or message $w$ into the source model, $M_o$. This creates a modified, watermarked model $M_w$, which is then released. To audit a suspicious model $M_s$, the model owner could verify whether the predefined signal $w$ can be extracted from it.
\end{definition}

Despite its effectiveness, LLM watermarking suffers from several practical limitations~\cite{li2023protecting, pasquini2025llmmap}. \textbf{(1)} The embedding process inevitably modifies the model's parameters, which can degrade its performance on primary tasks. \textbf{(2)} This process often requires substantial additional computational overhead, such as fine-tuning the model to embed the watermark. \textbf{(3)} Most critically, model watermarking is not applicable to models that have already been trained and released, as it must be performed beforehand. These limitations motivate the development of LLM fingerprinting as an alternative auditing solution~\cite{zeng2024huref}.

\partitle{LLM Fingerprinting} LLM fingerprinting is a post-hoc approach that does not require modifying the model. Instead, it extracts a set of inherent yet distinctive characteristics that collectively serve as the model's ``fingerprint''~\cite{pasquini2025llmmap, zhang2025reef, zeng2024huref}. The core idea is that any model derived from $M_o$ will preserve a statistically significant portion of this fingerprint. The definition is as follows.


\begin{definition}[LLM Fingerprinting]
LLM fingerprinting involves first identifying and extracting a set of inherent yet distinctive characteristics $F_o$ from the source model $M_o$, which collectively serve as its fingerprint. To audit a suspicious model $M_s$, the model owner could verify whether a fingerprint similar to $F_o$ can be extracted from $M_s$.
\end{definition}

The primary advantage of LLM fingerprinting lies in its \emph{non-invasive} nature. Since it does not alter the model, it introduces no performance degradation. The computational overhead is typically confined, which is often less intensive than the fine-tuning required for watermarking. Consequently, fingerprinting offers a more flexible and widely applicable paradigm for copyright auditing, especially for models that are already in the public domain~\cite{yao2023removalnet, pasquini2025llmmap, shao2025fit}. As such, in this paper, we primarily focus on LLM fingerprinting techniques.

\section{Principle of LLM Fingerprinting}

In this section, we present a comprehensive and unified formulation, including the framework, objectives, and taxonomy, of LLM fingerprinting. 

\subsection{LLM Fingerprinting Framework}

Generally, an LLM fingerprinting scheme can be defined by a formal, two-stage protocol. This protocol is built upon two fundamental functions: one is \emph{the extraction function} $\text{Extract}(\cdot)$ for extracting a signature, and the other is \emph{the fingerprint verification function} $\text{Verify}(\cdot)$ for verifying it against a suspicious model. 

\begin{itemize}[leftmargin=*]
    \item $\text{Extract}(M_o)\rightarrow F_o$: This function serves as the feature extractor. It takes the source model $M_o$ as input and produces a fingerprint $F_o$ of $M_o$. The fingerprint $F_o$ is not the model itself, but a compressed representation of its most fundamental and unique characteristics. 
    \item $\text{Verify}(M_s, F_o)\rightarrow 0/1$: The fingerprint verification takes the suspicious model $M_s$ and the source model's fingerprint $F_o$ as input. The goal of this function is to determine whether a statistically similar fingerprint can be extracted from $M_s$. It outputs $1$ if and only if a similar fingerprint is detected in $M_s$, and $0$ otherwise.
\end{itemize}

The framework hinges on the careful design of these two functions. The $\text{Extract}(\cdot)$ function needs to isolate a signal that is sensitive to a model's provenance. The $\text{Verify}(\cdot)$ function then measures this signal to make a final, decisive judgment. Together, they provide a systematic methodology for navigating the challenge of LLM copyright auditing.

\subsection{Design Objectives}
\label{sec:objectives}

For an LLM fingerprinting method to be reliable and practical, it must satisfy four fundamental design objectives. These objectives collectively characterize the essential principles of a successful system for LLM copyright auditing.

\begin{itemize}[leftmargin=*]
    \item \textbf{Effectiveness and Robustness:} The method should successfully identify a derivative model. If a suspicious model $M_s$ was indeed developed from the source model $M_o$, the fingerprinting process should return a positive match. This is the primary goal of the LLM copyright auditing.
    \item \textbf{Uniqueness:} The fingerprint must be distinctive to the model's lineage and should not falsely match models developed independently. Ensuring a low false positive rate is essential to avoid wrongful accusations. 
    \item \textbf{Discriminability:} To achieve both high effectiveness and uniqueness, the difference between fingerprints of derivative and independent models should be maximized, enabling reliable differentiation.
    \item \textbf{Efficiency:} The fingerprinting process must be computationally lightweight. A practical auditing method should incur only modest time and resource overhead, enabling fast and scalable analysis of suspicious models. 
\end{itemize}

\subsection{Taxonomy of LLM Fingerprinting}

In this paper, we first classify existing LLM fingerprinting methods based on a practical criterion: the level of access an auditor has to the suspicious model, $M_s$. This primary distinction is crucial because the information accessible in white-box and black-box scenarios differs fundamentally, leading to significant disparities in their technical paradigms and evaluation methods. Based on this, existing techniques fall into one of two categories: white-box~\cite{zeng2024huref, zhang2025reef} or black-box LLM fingerprinting~\cite{pasquini2025llmmap, jin2024proflingo}.

\begin{itemize}
    \item \textbf{White-box LLM Fingerprinting:} This category assumes the auditor has full access to the suspicious model's internal components, including its architecture and parameters. With this level of access, fingerprints can be extracted by directly analyzing the model's static weights or its dynamic internal states.
    \item \textbf{Black-box LLM Fingerprinting:} This category assumes the auditor can only interact with the suspicious model through an API. Access is restricted to submitting queries and observing the outputs. In this category, the fingerprint can be inferred exclusively from the model's input-output behavior, without any knowledge of its internal structure.
\end{itemize}

This distinction in access level fundamentally shapes the practical applicability of each category. White-box fingerprinting methods are best suited for auditing open-source models, where auditors can directly download and examine model parameters. In contrast, black-box fingerprinting methods apply to a broader set of scenarios, including the auditing of proprietary commercial models accessed via APIs, as well as cases of model theft in which an adversary deploys a stolen model as a closed service. 

Building upon this foundational division, we will further provide a detailed sub-classification for both white-box and black-box approaches. We classify white-box methods into three different subcategories, \emph{static}, \emph{forward-pass}, and \emph{backward-pass fingerprinting}, and black-box methods into two different subcategories, \emph{untargeted} and \emph{targeted fingerprinting}. Detailed introductions are in Section \ref{sec:white-box}\&\ref{sec:black-box}. A summary of existing approaches is provided in Appendix~\ref{sec:summary}.

\section{White-box LLM Fingerprinting}
\label{sec:white-box}

\begin{figure}
    \centering
    \includegraphics[width=0.99\linewidth]{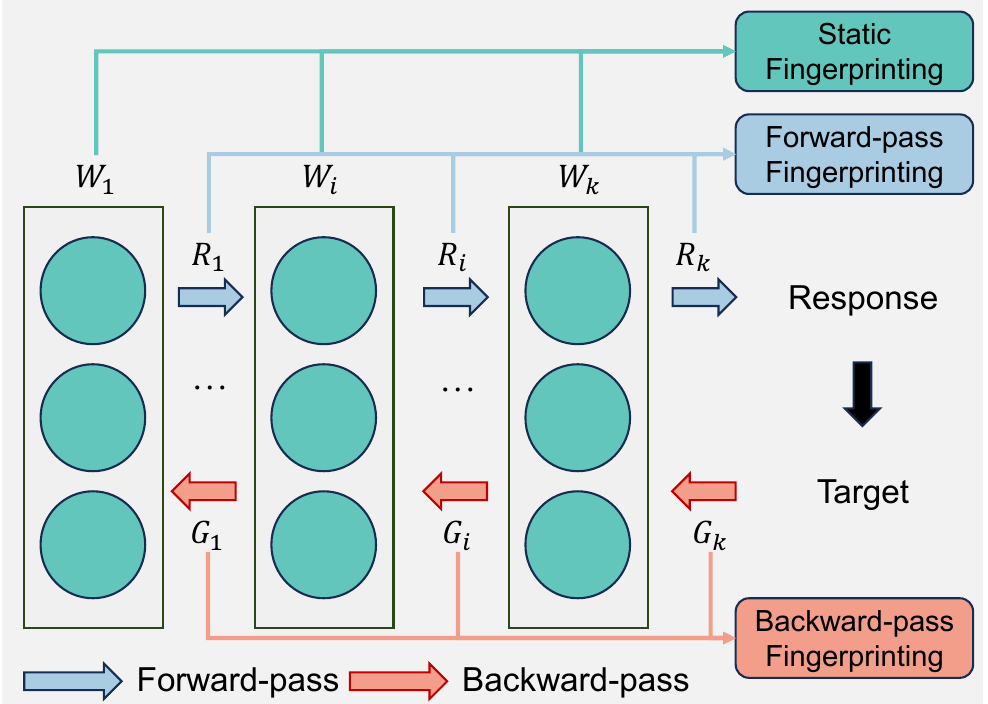}
    \caption{The taxonomy of white-box LLM fingerprinting.}
    \label{fig:white-box}
    \vspace{-0.5em}
\end{figure}

White-box LLM fingerprinting methods operate under the assumption of full access to a model's internal architecture and parameters. We classify these methods into three different categories, \emph{static}, \emph{forward-pass}, and \emph{backward-pass fingerprinting}. This taxonomy is determined by whether the features are derived from the model in a static state, during a forward pass, or during a backward pass, as Figure~\ref{fig:white-box}. 

\subsection{Static Fingerprinting}

Static fingerprinting methods~\cite{li2021modeldiff, zheng2022dnn, zeng2024huref, yoon2025intrinsic} are the most straightforward form of white-box auditing. They operate directly on the model's learned parameters $\{W_1, \dots, W_k\}$, treating the static weights themselves as the primary source of identifying information.

\partitle{Architecture-agnostic Fingerprinting} Some static fingerprinting approaches treat a model's parameters as a single undifferentiated tensor of numerical values and extract fingerprints from all or part of the model weights. A simple method, WeightCompare~\cite{li2021modeldiff}, measures model similarity by computing a score based on the number of layers with identical structures and weights across two models. Beyond such direct comparisons, the more advanced technique aims to generate fingerprints that are more robust and efficient. Zheng et al.~\cite{zheng2022dnn} proposed one such method, which produces a non-repudiable fingerprint by projecting front-layer weights onto a random space defined by the model owner's unique identity. 
These methods are architecture-agnostic and do not depend on any specific model design.

\partitle{Transformer-specific Fingerprinting} Currently, modern LLMs are mostly based on the Transformer architecture~\cite{vaswani2017attention}. Based on this understanding, a more targeted method tailors the fingerprinting process to this specific structure. Instead of viewing parameters as a generic block of data, these methods selectively extract features from key architectural components (\ie, query, key, value, and output matrices) of transformers. Zeng et al.~\cite{zeng2024huref} proposed HuRef, which constructs invariant terms by combining weight matrices from attention and feed-forward layers, designing them to be resilient against manipulations like parameter permutation. More straightforwardly, Yoon et al.~\cite{yoon2025intrinsic} discover that the standard deviation distributions of attention matrices across layers could form an intrinsic fingerprint. Recently, Zhang proposed MDIR~\cite{zhang2025matrix}, a method using matrix analysis to detect and reconstruct weight relationships. MDIR leverages polar decomposition to compare the orthogonal parts of weight matrices and can identify the exact transformation between them, providing a statistical p-value for the detected similarity using Large Deviation Theory.


\subsection{Forward-pass Fingerprinting}

Forward-pass fingerprinting methods~\cite{chen2022copy, zhang2024easydetector, zhang2025reef} leverage the dynamic internal states of an LLM as it processes input data. The fundamental idea is that during a forward pass, a model generates a vast number of intermediate representations $\{R_1, \dots, R_k\}$ ($k$ is the number of layers), which are shaped by its unique architecture and learned parameters. These high-dimensional features can serve as a distinctive fingerprint to characterize the model's behavior and origins. Unlike static methods that analyze model weights in isolation, forward-pass techniques capture the functional properties of a model in action.


Recent works have proposed various instantiations of forward-pass fingerprinting. A pioneering work named DeepJudge~\cite{chen2022copy} introduced a comprehensive testing framework using multiple distance metrics on intermediate model outputs. It generates specialized test cases, such as adversarial examples or inputs that could trigger corner-case neuron activations, to amplify differences, which are then quantified using direct distance measures. Following DeepJudge, EasyDetector~\cite{zhang2024easydetector} employs an indirect comparison method by training a simple linear classifier (also called ``linear probe'') on the source model's representations. The probe's classification confidence on a suspect model then serves as the similarity score. To enhance robustness, REEF~\cite{zhang2025reef} uses Centered Kernel Alignment to compare the geometry of different models' representations.

\subsection{Backward-pass Fingerprinting}

Backward-pass fingerprinting methods represent a more advanced form of dynamic analysis, moving beyond the intermediate features of a forward pass. The core idea is to use the gradients $\{G_1, \dots, G_k\}$ (or the features of the gradients) generated during backpropagation as the model's fingerprint. Gradients reveal how a model's parameters would be updated in response to small input perturbations, offering a fine-grained signature of its learned optimization landscape and internal sensitivities. This solution may be able to capture deeper functional characteristics than forward-pass features alone, as it reflects the model's learning dynamics.

To the best of our knowledge, the only existing work on backward-pass fingerprinting for LLMs is TensorGuard~\cite{wu2025gradient}. Its core idea is to characterize a model using statistical features of gradients (\eg, mean, standard deviation, norm). To generate a stable and robust signature, TensorGuard perturbs the model by injecting controlled random noise into internal tensors during a forward pass, which induces consistent gradient responses captured via backpropagation. Statistical features are then extracted from these gradients, and principal component analysis is applied for dimensionality reduction to enhance efficiency. Finally, similarity is measured by computing the Euclidean distance between the resulting fingerprint vectors. 


\section{Black-box LLM Fingerprinting}
\label{sec:black-box}

Black-box LLM fingerprinting methods operate under the constraint that the auditor has no access to the suspect model's internal components, such as its architecture or parameters. Instead, the auditor can only interact with the model through its public API by issuing queries and analyzing the corresponding responses. The central challenge in this setting is to derive a unique and reliable fingerprint solely from observable input–output behavior. In our taxonomy, we further classify black-box fingerprinting into two categories: untargeted and targeted, depending on whether the method relies on detecting developer-predefined input–output pairs.


\subsection{Untargeted Fingerprinting}

Untargeted fingerprinting~\cite{yang2024fingerprint, iourovitski2024hide, pasquini2025llmmap} represents a direct and intuitive approach to black-box LLM copyright auditing. Several preliminary studies observed that the outputs generated by different LLMs are often distinguishable, exhibiting unique stylistic or behavioral ``idiosyncrasies''~\cite{sun2025idiosyncrasies, bitton2025detecting, suzuki2025natural}. Inspired by this observation, untargeted fingerprinting typically operates by first selecting a set of input queries and taking the (features of) corresponding output responses as fingerprints. A high degree of similarity between the responses of different LLMs could indicate that one is a derivative of the other.

\begin{figure}[t]
    \centering
    \includegraphics[width=0.99\linewidth]{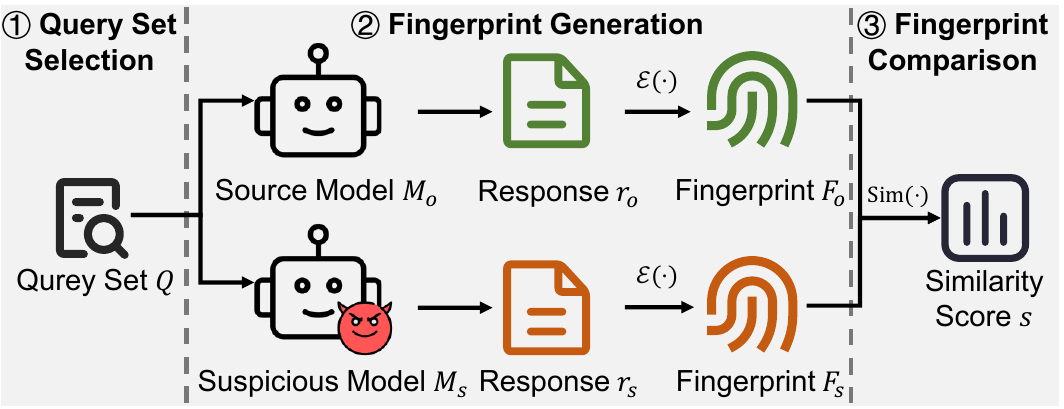}
    \caption{The paradigm of untargeted LLM fingerprinting.}
    \label{fig:untargeted}
    \vspace{-0.5em}
\end{figure}

\partitle{Paradigm} The paradigm can be formalized as a three-stage pipeline, as shown in Figure~\ref{fig:untargeted}. Given an source model $M_o$ and a suspicious model $M_s$, the process is as follows:
\begin{enumerate}
    \item \textbf{Query Set Selection}: A dataset of input queries, denoted as $Q=\{q_1, q_2, \dots, q_n\}$, is chosen to probe the model's response and behavior. 
    \item \textbf{Fingerprint Generation}: In this stage, the fingerprint is generated by capturing the model's responses to the selected queries. This process involves first querying the models $M_o$ and $M_s$ with the query set $Q$ to obtain the raw output responses. Subsequently, a feature extraction function $\mathcal{E}(\cdot)$ is applied to transform these responses into structured fingerprints, $F_o$ and $F_S$, as follows:
    \begin{equation}
        F_o=\mathcal{E}(M_o(Q));\enspace F_s=\mathcal{E}(M_s(Q)).
    \end{equation}
    \item \textbf{Fingerprint Comparison}: A specific similarity metric or function, $\text{Sim}(\cdot, \cdot)$ is used to compare the fingerprints $F_o$ and $F_s$. This yields a final judgment score $s$, as shown in Eq.~(\ref{eq:untarget_sim}). This score is typically compared against a predetermined threshold $\tau$ to make the decision.
    \begin{equation}
        \label{eq:untarget_sim}
        s = \text{Sim}(F_s, F_o).
    \end{equation}
\end{enumerate}

The core innovations and primary distinctions among existing untargeted fingerprinting techniques lie in the specific designs of the query set $Q$, the extraction function $\mathcal{E}(\cdot)$, and the similarity function $\text{Sim}(\cdot, \cdot)$.

\partitle{Choice of the Query Set $Q$} 
The choice of the query set $Q$ is critical as it determines the nature of the behavioral signals elicited from the models. Existing works have explored several strategies for selecting or generating these queries: 
\begin{itemize}
    \item \textbf{Manually Constructed Query Sets}: This method uses queries that are either strategically handcrafted or sampled from existing datasets. Methods like LLMmap~\cite{pasquini2025llmmap} and DuFFin~\cite{yan2025duffin} manually curate queries that target specific model behaviors, such as alignment and knowledge across various domains. Other methods like model equality testing (MET)~\cite{gao2025model} and Rank-Based Uniformity Test (RUT)~\cite{zhu2025auditing} instead use prompts from general corpora like Wikipedia or user-chatbot conversations to mimic natural user interaction patterns.
    \item \textbf{Automatically Generated Query Sets}: To discover more discriminative queries, some works also explore automating the generation process. For instance, Hide-and-Seek~\cite{iourovitski2024hide} employs an evolutionary algorithm where an LLM iteratively generates and refines prompts to maximize response differences among various models. Alternatively, Kurian et al.~\cite{kurian2025attacks} employ reinforcement learning to automatically select the optimal query subset from a larger, pre-defined pool.
    \item \textbf{Query Set Exploring Specific LLM Capabilities}: A third method focuses on fingerprinting a single, complex capability of an LLM. For instance, CoTSRF~\cite{ren2025cotsrf} uses a query set of reasoning questions paired with a Chain-of-Thought (CoT) trigger to specifically probe and capture the model's unique logical reasoning patterns.
\end{itemize}

\partitle{Method of the Extraction Function $\mathcal{E}(\cdot)$} Once the model’s responses are collected, the extraction function $\mathcal{E}(\cdot)$ transforms them into a structured and compact fingerprint. Different fingerprinting methods vary in their level of abstraction. 
\begin{itemize}[leftmargin=*]
    \item \textbf{Using Output Logits or Probabilities}: The most direct methods operate on a model's output logits or probabilities, such as reconstructing a unique vector space from the probability distribution~\cite{yang2024fingerprint} or computing a scalar log-rank score for each response~\cite{zhu2025auditing}.
    \item \textbf{Leveraging Lexical and Syntactic Features}: Other methods analyze the generated text for its lexical and syntactic features~\cite{bhardwaj2025invisible}. For instance, McGovern et al.~\cite{mcgovern2025your} show that character, word, and part-of-speech (POS) n-grams could serve as a robust fingerprint.
    \item \textbf{Employing Neural Networks}: The most prevalent techniques employ neural networks to learn a feature embedding. Methods including LLMmap~\cite{pasquini2025llmmap}, CoTSRF~\cite{ren2025cotsrf}, DuFFin~\cite{yan2025duffin}, and Kurian et al.'s method~\cite{kurian2025attacks} all leverage Transformer-based models to encode responses into a dense fingerprint vector that captures the model's features.
\end{itemize}

\partitle{Design of the Similarity Function $\rm{Sim}(\cdot, \cdot)$}
Finally, the similarity function $\rm{Sim}(\cdot, \cdot)$ compares the extracted fingerprints to decide model identity, with existing methods ranging from simple distance measures to formal statistical tests. The most straightforward approaches \emph{compute a distance or similarity score directly in the feature space}, using metrics such as cosine similarity~\cite{pasquini2025llmmap, yan2025duffin}, Hamming distance~\cite{yan2025duffin}, or distributional KL-divergence~\cite{ren2025cotsrf}. More advanced approaches \emph{formulate the task as a classification problem}, training a model (\eg, a GradientBoost classifier~\cite{mcgovern2025your} or a dedicated Transformer network~\cite{pasquini2025llmmap, bhardwaj2025invisible}) to map fingerprints back to their source LLMs. At the most formal level, fingerprint comparison is treated as \emph{a statistical two-sample hypothesis test}. For example, Gao et al.\cite{gao2025model} employ a Maximum Mean Discrepancy (MMD) test, while RUT\cite{zhu2025auditing} applies a Cramér–von Mises test to rigorously determine whether two sets of fingerprints are drawn from the same distribution.

\subsection{Targeted Fingerprinting}
In contrast to untargeted fingerprinting, targeted fingerprinting seeks to construct a specific set of query–response pairs that are unique to a source model and its derivatives~\cite{gubri2024trap, jin2024proflingo, xu2025rap, tsai2025rofl}. The key idea is that these pairs, collectively forming the fingerprint, will be consistently reproduced only by models in the source lineage, whereas independently trained models will fail to generate the corresponding target responses. The essential distinction between untargeted and targeted approaches lies in \emph{whether fingerprint verification relies on predefined query–response pairs}. Notably, targeted methods yield an \emph{asymmetric} fingerprint: for a given targeted method $\mathcal{A}$, the outcome of $\mathcal{A}(M_o, M_s)$ is not necessarily equal to $\mathcal{A}(M_s, M_o)$.


\begin{figure}[t]
    \centering
    \includegraphics[width=0.99\linewidth]{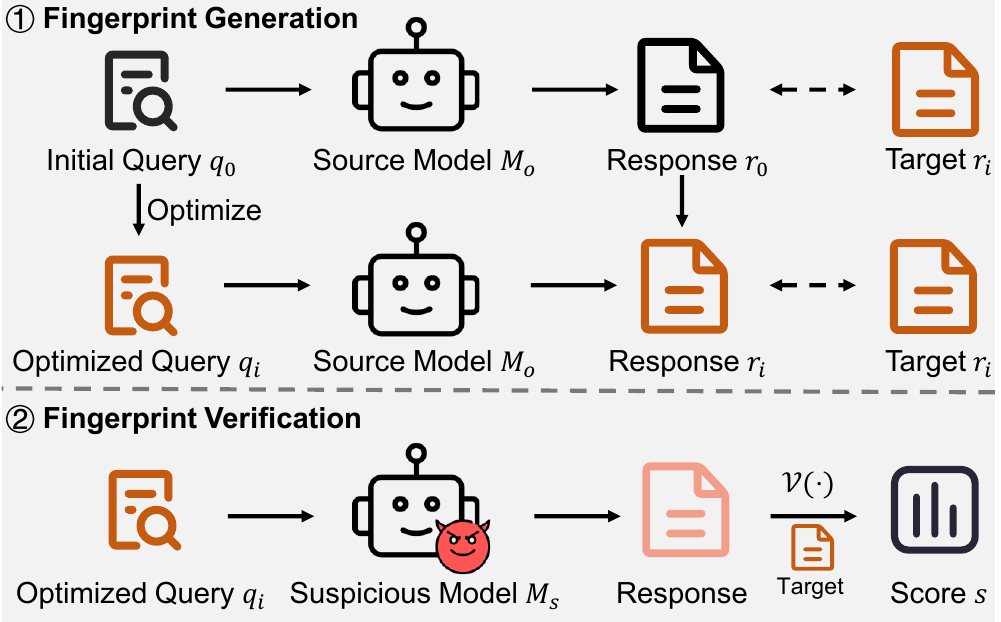}
    \caption{The paradigm of targeted LLM fingerprinting.}
    \label{fig:targeted}
    \vspace{-0.5em}
\end{figure}

\partitle{Paradigm} The paradigm of targeted fingerprinting methods consists of two stages, as depicted in Figure~\ref{fig:targeted}:
\begin{enumerate}
    \item \textbf{Fingerprint Generation}: In this stage, a set of unique query-response pairs, $F_o=\{(q_i, r_i)\}_{i=1}^n$ is generated from the source model $M_o$. This is typically achieved through a search or optimization process that seeks to find queries $q_i$ that elicit highly specific and unique responses $r_i$ from $M_o$. The generation process ensures that for every pair in the fingerprint, the condition in Eq.~(\ref{eq:target_generation}) holds:
  \begin{equation}
  \label{eq:target_generation}
      r_i=M_o(q_i), \forall (q_i,r_i)\in F_o.
  \end{equation}
  \item \textbf{Fingerprint Verification}: To audit a suspicious model $M_s$, the queries from the generated fingerprint, $Q_o=\{q_1, \dots, q_n\}$, are submitted to $M_s$. The verification function then checks whether the responses from $M_s$ match the predefined responses in $F_o$. If a significant number of responses match, the suspicious model is considered a likely derivative. This can be formalized by a verification score, $\mathcal{V}(M_s, F_o)$, calculated as the proportion of matching pairs:
  \begin{equation}
      \mathcal{V}(M_s, F_o)=\frac{1}{n}\sum_{i=1}^n\mathbb{I}(\text{Sim}(M_s(q_i), r_i) \geq \mu),
  \end{equation}
  where $\text{Sim}(\cdot,\cdot)$ is a similarity metric function and $\mu$ is a predefined similarity threshold.  The suspicious model $M_s$ is identified as a derivative of $M_o$ if this score exceeds another predetermined threshold $\tau$.
\end{enumerate}

The primary innovation within targeted fingerprinting lies in methods for generating unique query-response pairs, often using gradient-based optimization inspired by adversarial attacks~\cite{zou2023universal}. Methods like TRAP~\cite{gubri2024trap} and ProFLingo~\cite{jin2024proflingo} craft an adversarial prompt component to force a source model to produce a specific, pre-determined response, such as a designated number or a factually incorrect answer. To enhance the robustness of this fingerprint across an entire model lineage, RAP-SM~\cite{xu2025rap} extends this technique by using shadow models (\ie, known derivatives of the source model) for joint optimization. Furthermore, a distinct method, RoFL~\cite{tsai2025rofl}, generates fingerprints without a pre-defined target response. Instead, it first gets a response by feeding an unlikely token sequence to the source model. Then RoFL optimizes the prompt to consistently re-elicit the same response across model variants.

\section{\bench: Benchmarking LLM Fingerprinting in the Wild}

While numerous LLM fingerprinting techniques have been proposed, they are often evaluated under disparate settings, making direct and fair comparison a significant challenge. To date, the field has still lacked a unified and standardized benchmark for assessing the effectiveness and reliability of these methods. To address this gap, we introduce \bench, a benchmark designed to facilitate a systematic evaluation of LLM fingerprinting techniques against a curated representative set of realistic model modifications and deployment scenarios. Besides, \bench provides extensible and user-friendly interfaces, enabling the evaluation of new fingerprinting methods and additional post-development techniques.


In this section, we first summarize key techniques commonly used in real-world LLM applications that may affect the effectiveness of fingerprinting methods. We then describe the design and components of \bench, which integrates a representative selection of these techniques to construct practical auditing scenarios. Finally, we conduct an empirical evaluation of several SOTA fingerprinting methods on \bench and highlight our key findings.

\subsection{Techniques affecting LLM Fingerprinting}
\label{sec:technique}

After an LLM is pre-trained, it often undergoes a variety of post-development modifications before being deployed in a real-world application. These techniques, designed to adapt or optimize the model, can significantly alter its parameters and behavior. Consequently, the stability and detectability of its fingerprint may be impacted. We broadly classify these influential techniques into two categories: \emph{parameter-altering} and \emph{parameter-independent techniques}.

\partitle{Parameter-altering Techniques} These techniques refer to those that permanently modify the model's parameters. These methods are typically applied to adapt a general-purpose pre-trained model for a specific downstream task or for better efficiency. By directly changing the model's weights, these techniques can fundamentally alter an LLM's intrinsic characteristics. Representative examples include:

\begin{itemize}
    \item \textbf{Fine-tuning}. It is the process of further training a pre-trained model on a smaller and task-specific dataset (\ie, personalization dataset) to specialize its capabilities~\cite{hu2022lora}.  
    \item \textbf{Quantization}. It is a compression technique that reduces the precision of model weights (\eg, from 32 bits to 8 bits), thereby making the model smaller and faster~\cite{jin2024comprehensive}.  
    \item \textbf{Distillation}. It is used to train a smaller ``student'' model for mimicing a larger ``teacher'' model, transferring knowledge into a more compact form~\cite{yang2024survey}.
    \item \textbf{Model Merge}. It is a technique that combines the weights of two or more different models to create a new model inheriting capabilities from its predecessors~\cite{goddard2024arcee}.  
    
\end{itemize}

\partitle{Parameter-independent Techniques} These techniques influence the outputs of LLMs during inference without altering their underlying weights. They guide or constrain the generation process to improve performance, mitigate hallucinations, align responses with human values, or enhance output diversity. Representative examples include:


\begin{itemize}
    \item \textbf{System Prompts}: System prompts are high-level instructions provided to the model alongside the user's query to set a specific context or set of rules for the interaction~\cite{giray2023prompt}.
    \item \textbf{Sampling Strategies}: Sampling strategies are algorithms that control how the next token is selected from the model's output probability distribution. Common techniques include Temperature scaling, Top-k, and Top-p sampling, which adjust the randomness and diversity of the generated text~\cite{zhang2023survey}.
    \item \textbf{Retrieval-Augmented Generation (RAG)}: RAG enhances model responses by retrieving relevant information from an external knowledge base and supplying it as additional context during generation~\cite{lewis2020retrieval}.  
\end{itemize}



\subsection{Construction of \bench}
\label{sec:overview}

Based on the above summary, \bench consists of three core components: a collection of pre-trained models and two categories of post-development techniques, which are integrated to enable diverse evaluation scenarios.

\partitle{Pre-trained Models} \bench is built upon 7 mainstream pre-trained foundation models, including variants of Qwen~\cite{qwen2.5}, Llama~\cite{grattafiori2024llama3, touvron2023llama2}, Mistral~\cite{mistral}, Gemma~\cite{gemma2}, and TinyLlama~\cite{zhang2024tinyllama}. These models were selected for their popularity and advanced capabilities, with parameter counts ranging from 1.1B to 14B. Crucially, they all have a widespread community with a large number of publicly available derivative models on platforms like Hugging Face, making them ideal for studying copyright lineage in the wild.

\partitle{Implementation of Parameter-altering Techniques} To simulate how adversaries adapt pre-trained models, \bench includes derivative models created using various widely-used parameter-altering techniques, including: \textbf{(1)} instruction tuning (IT), \textbf{(2)} general-purpose fine-tuning (FT), \textbf{(3)} parameter-efficient fine-tuning (PEFT), \textbf{(4)} quantization (QZ), \textbf{(5)} model merging (MM), and \textbf{(6)} distillation (DT). For each of the seven pre-trained models, we have collected no fewer than seven corresponding derivative models from Hugging Face that have been modified using one or more of these methods.

\partitle{Implementation of Parameter-independent Techniques} \bench also evaluates the robustness of fingerprinting methods against common runtime parameter-independent techniques. This includes: \textbf{(1)} system prompts, covering general-purpose prompts (GP), role-playing prompts (RP), and chain-of-thought (CoT) prompts; \textbf{(2)} sampling strategies (SS), with configurations for greedy decoding, high-randomness, low-randomness, and moderate settings; \textbf{(3)} LLM with RAG; and \textbf{(4)} adversarial manipulation (ADV), which considers two scenarios: rewriting the input query with a smaller LLM and perturbing the output logits with random noise.


\partitle{Organizing the Model Instances into Triplets} In total, \bench includes \textbf{149} distinct model instances. To operationalize our evaluation, we structure \bench to directly address the copyright auditing problem formalized in Definition~\ref{def:audit}. Specifically, we construct a comprehensive dataset in which each sample is represented as a triplet $(M_o^i, M_s^i, y^i)$. Here, $M_o^i$ denotes the source model, selected from a pool of 13 base models, including both pre-trained (PT) and instruction-tuned (IT) variants. This design captures realistic auditing scenarios, where developers may seek to protect either their foundational pre-trained models or their specialized instruction-following derivatives. Besides, the suspicious model $M_s^i$ is drawn from the complete set of 149 model instances, encompassing all base models and their derivatives. The label $y^i \in \{0,1\}$ serves as the ground truth for the auditing task, where $y^i = 1$ if and only if $M_s^i$ is a known derivative of $M_o^i$, and $y^i = 0$ otherwise. This formulation results in a total of $13 \times 149 = 1937$ evaluation samples within the designed \bench.


A comparison between existing studies and \bench is presented in Table~\ref{tab:comparison_bench}. Notably, given the complexity and rapid evolution of LLM post-development techniques, it is difficult for this benchmark to be fully exhaustive. Nonetheless, by focusing on a representative set of widely adopted techniques, \bench strikes a balance between comprehensiveness and efficiency. This design enables it to serve as a practical tool for assessing the effectiveness and robustness of existing methods. Further details are provided in Appendix~\ref{sec:details_bench}.



\begin{table}[t!]
\centering
\tabcolsep=0.9mm
\renewcommand{\arraystretch}{1.1}
\caption{Comparison of \bench with the evaluation settings of representative existing studies. \cmark\xspace indicates the technique is included, while \xmark\xspace indicates it is not.}
\label{tab:comparison_bench}
\scalebox{0.72}{
\begin{tabular}{l c c c c c c c c c c c c c}
\toprule
\cmidrule(lr){2-5} \cmidrule(lr){6-9}
\textbf{Method} & PT & IT & FT & PEFT & QZ & MM & DT & GP & RP & CoT & SS & RAG & ADV\\
\midrule
HuRef~\cite{zeng2024huref} & \cmark & \cmark & \cmark & \cmark & \xmark & \cmark & \xmark & \xmark & \xmark & \xmark & \xmark & \xmark & \cmark\\
REEF~\cite{zhang2025reef} & \cmark & \cmark & \cmark & \xmark & \xmark & \cmark & \xmark & \xmark & \xmark & \xmark & \xmark & \xmark & \cmark\\
PDF~\cite{yoon2025intrinsic}  & \cmark & \cmark & \cmark & \xmark & \xmark & \xmark & \xmark & \xmark & \xmark & \xmark & \xmark & \xmark & \xmark \\
LLMmap~\cite{pasquini2025llmmap} & \xmark & \cmark & \cmark & \xmark & \xmark & \xmark & \xmark & \cmark & \cmark & \cmark & \cmark & \cmark & \xmark\\
MET~\cite{gao2025model} & \xmark & \cmark & \cmark & \xmark & \cmark & \xmark & \xmark & \xmark & \xmark & \xmark & \xmark & \xmark & \xmark \\
TRAP~\cite{gubri2024trap} & \xmark & \cmark & \xmark & \xmark & \xmark & \xmark & \xmark & \cmark & \cmark & \xmark & \cmark & \xmark & \xmark\\
\midrule
\textbf{\bench} & \cmark & \cmark & \cmark & \cmark & \cmark & \cmark & \cmark & \cmark & \cmark & \cmark & \cmark & \cmark & \cmark \\ 
\bottomrule
\end{tabular}
}
\end{table}

\begin{table*}[t]
\centering
\tabcolsep=1.1mm
\renewcommand{\arraystretch}{1.2}
\caption{Overall performance comparison of different LLM fingerprinting methods.}
\label{tab:overall}
\scalebox{0.82}{
\begin{tabular}{l c c c c c c c c}
\toprule
 & \multicolumn{4}{c}{\textbf{White-box}} & \multicolumn{4}{c}{\textbf{Black-box}} \\
\cmidrule(lr){2-5} \cmidrule(lr){6-9}
\textbf{Metric} & HuRef & PDF & REEF & Gradient & LLMmap & MET & SEF & TRAP \\
\midrule
\textbf{AUC} $\uparrow$ & 0.994($\pm$0.000) & \textbf{0.995($\pm$0.000)} & 0.896($\pm$0.021) & 0.798($\pm$0.008) & 0.632($\pm$0.009) & 0.654($\pm$0.009) & 0.581($\pm$0.008) & \textbf{0.712}($\pm$0.027)\\
\textbf{pAUC} $\uparrow$ & 0.971($\pm$0.000) & \textbf{0.996($\pm$0.000)} & 0.832($\pm$0.019) & 0.708($\pm$0.008) & 0.635($\pm$0.009) & \textbf{0.672}($\pm$0.005) & 0.646($\pm$0.004) & 0.625($\pm$0.006)\\
\textbf{MD} $\uparrow$ & \textbf{3.154($\pm$0.000)} & 1.741($\pm$0.000) & 2.091($\pm$0.058) & 0.672($\pm$0.031) & 0.521($\pm$0.021) & \textbf{1.681}($\pm$0.051) & 0.172($\pm$0.014) & 1.382($\pm$0.064)\\
\textbf{ACC} $\uparrow$ & 0.992($\pm$0.000) & \textbf{0.996($\pm$0.000)} & 0.892($\pm$0.021) & 0.812($\pm$0.032) & 0.844($\pm$0.018) & \textbf{0.892}($\pm$0.005) & 0.879($\pm$0.004) & 0.820($\pm$0.037)\\
\textbf{TPR@1\%FPR} $\uparrow$ & 0.943($\pm$0.000) & \textbf{0.992($\pm$0.000)} & 0.634($\pm$0.021) & 0.374($\pm$0.031) & 0.253($\pm$0.020) & \textbf{0.356}($\pm$0.007) & 0.282($\pm$0.012) & 0.227($\pm$0.020)\\
\bottomrule
\end{tabular}
}
\end{table*}

\subsection{Experimental Settings}
\label{sec:settings}

\firstpartitle{Evaluated LLM Fingerprinting Methods} To provide a comprehensive evaluation, we select, design, and implement 8 representative fingerprinting methods that span the full taxonomy of our proposed framework. \textbf{(1)} For \emph{static fingerprinting}, we evaluate two methods: HuRef~\cite{zeng2024huref} and Parameter Distribution Fingerprint (PDF)~\cite{yoon2025intrinsic}. \textbf{(2)} For \emph{forward-pass fingerprinting}, we include REEF~\cite{zhang2025reef}. \textbf{(3)} In the \emph{backward-pass fingerprinting}, we implement a method named Gradient, which is inspired by the approach of TensorGuard~\cite{wu2025gradient} and incorporates several improvements. \textbf{(4)} For \emph{untargeted fingerprinting}, we evaluate three black-box methods: LLMmap~\cite{pasquini2025llmmap}, Model Equality Testing (MET)~\cite{gao2025model}, and our proposed simple baseline, Sentence Embedding Fingerprinting (SEF), inspired by \cite{sun2025idiosyncrasies}. In general, SEF uses the averaged sentence embeddings of model responses as the fingerprint (see Appendix~\ref{sec:sef} for details). \textbf{(5)} Finally, for \emph{targeted fingerprinting}, we select TRAP~\cite{gubri2024trap}, which uses adversarial suffixes to generate unique query-response pairs. In particular, for all methods, we conduct at least 5 trials and report both the mean and the standard deviation.


\partitle{Metrics} To evaluate the fingerprinting methods against the primary design objectives outlined in Section~\ref{sec:objectives}, we employ three primary evaluation metrics, defined as follows:

\begin{itemize}
    \item \textbf{Area Under the ROC Curve (AUC)}: This metric provides a comprehensive measure of a method's ability to distinguish between derivative and independent models, reflecting its overall \textbf{Effectiveness and Robustness}. A higher AUC indicates better performance.
    \item \textbf{Partial AUC with FPR$\in[0, 0.05]$ (pAUC)}: This metric evaluates performance in a low False Positive Rate (FPR) range (\ie, $[0, 0.05]$). It highlights a method's \textbf{Uniqueness} in avoiding false positives. Notably, this metric is standardized to $[0,1]$ in our experiments.
    \item \textbf{Mahalanobis Distance (MD)}: MD~\cite{xiang2008learning} is to measure the \textbf{Discriminability} of a fingerprinting method by calculating the distance between the fingerprint distributions of derivative and independent models. A larger distance signifies a clearer separation between model lineages. 
\end{itemize}

\begin{table*}[t]
\centering
\tabcolsep=1.5mm
\renewcommand{\arraystretch}{1.15}
\caption{Performance comparison of LLM fingerprinting methods across different source models. ``PT'' and ``IT'' refer to using the pre-trained models and instruction-tuned models as source models, respectively.}
\label{tab:source}
\scalebox{0.8}{
\begin{tabular}{lll ccccccccccccc}
\toprule
\multirow{2}{*}{\textbf{Type}} & \multirow{2}{*}{\textbf{Method}} & \multirow{2}{*}{\textbf{Metric}} & 
\multicolumn{2}{c}{\textbf{Qwen2.5-7B}} & 
\multicolumn{2}{c}{\textbf{Qwen2.5-14B}} & 
\multicolumn{2}{c}{\textbf{Llama3.1-8B}} & 
\multicolumn{2}{c}{\textbf{Mistral-7B}} & 
\multicolumn{2}{c}{\textbf{Gemma2-2B}} & 
\textbf{Tinyllama-1.1B} & 
\multicolumn{2}{c}{\textbf{Llama2-7B}} \\
\cmidrule(lr){4-5} \cmidrule(lr){6-7} \cmidrule(lr){8-9} \cmidrule(lr){10-11} \cmidrule(lr){12-13} \cmidrule(lr){14-14} \cmidrule(lr){15-16}
& & & \textbf{PT} & \textbf{IT} & 
\textbf{PT} & \textbf{IT} & \textbf{PT} & \textbf{IT} & 
\textbf{PT} & \textbf{IT} & 
\textbf{PT} & \textbf{IT} & 
\textbf{IT} & 
\textbf{PT} & \textbf{IT} \\
\midrule

\multirow{12}{*}{\textbf{White-box}} & \multirow{3}{*}{HuRef} 
    & AUC $\uparrow$      & \colorauc{1.000} & \colorauc{1.000} & \colorauc{1.000} & \colorauc{1.000} & \colorauc{1.000} & \colorauc{1.000} & \colorauc{1.000} & \colorauc{1.000} & \colorauc{1.000} & \colorauc{1.000} & \colorauc{0.937} & \colorauc{1.000} & \colorauc{1.000}\\
    & & pAUC $\uparrow$ & \colorauc{1.000} & \colorauc{1.000} & \colorauc{1.000} & \colorauc{1.000} & \colorauc{1.000} & \colorauc{1.000} & \colorauc{1.000} & \colorauc{1.000} & \colorauc{1.000} & \colorauc{1.000} & \colorauc{0.927} & \colorauc{1.000} & \colorauc{1.000}\\
    & & MD $\uparrow$       & \colormd{4.276} & \colormd{4.272} & \colormd{3.081} & \colormd{3.081} & \colormd{2.742} & \colormd{2.704} & \colormd{4.742} & \colormd{4.619} & \colormd{4.607} & \colormd{4.591} & \colormd{2.413} & \colormd{2.606} & \colormd{2.629}\\
\cmidrule{2-16}

& \multirow{3}{*}{PDF}
    & AUC $\uparrow$      & \colorauc{1.000} & \colorauc{1.000} & \colorauc{1.000} & \colorauc{1.000} & \colorauc{1.000} & \colorauc{1.000} & \colorauc{1.000} & \colorauc{1.000} & \colorauc{1.000} & \colorauc{1.000} & \colorauc{0.877} & \colorauc{1.000} & \colorauc{1.000}\\
    & & pAUC $\uparrow$ & \colorauc{1.000} & \colorauc{1.000} & \colorauc{1.000} & \colorauc{1.000} & \colorauc{1.000} & \colorauc{1.000} & \colorauc{1.000} & \colorauc{1.000} & \colorauc{1.000} & \colorauc{1.000} & \colorauc{0.927} & \colorauc{1.000} & \colorauc{1.000}\\
    & & MD $\uparrow$       & \colormd{1.753} & \colormd{1.756} & \colormd{2.926} & \colormd{2.925} & \colormd{1.369} & \colormd{1.367} & \colormd{1.226} & \colormd{1.226} & \colormd{13.995} & \colormd{14.164} & \colormd{1.442} & \colormd{2.182} & \colormd{2.182}\\
\cmidrule{2-16}

& \multirow{3}{*}{REEF}
    & AUC $\uparrow$   & \colorauc{0.824} & \colorauc{0.818} & \colorauc{0.934} & \colorauc{0.946} & \colorauc{0.996} & \colorauc{1.000} & \colorauc{1.000} & \colorauc{1.000} & \colorauc{0.942} & \colorauc{0.968} & \colorauc{0.770} & \colorauc{0.750} & \colorauc{0.750} \\
    & & pAUC $\uparrow$ & \colorauc{0.803} & \colorauc{0.763} & \colorauc{0.936} & \colorauc{0.919} & \colorauc{0.964} & \colorauc{0.997} & \colorauc{1.000} & \colorauc{1.000} & \colorauc{0.694} & \colorauc{0.858} & \colorauc{0.839} & \colorauc{0.872} & \colorauc{0.872}\\
    & & MD $\uparrow$       & \colormd{1.573} & \colormd{1.871} & \colormd{1.936} & \colormd{3.017} & \colormd{2.983} & \colormd{3.435} & \colormd{2.528} & \colormd{2.622} & \colormd{2.119} & \colormd{2.496} & \colormd{1.650} & \colormd{2.582} & \colormd{2.583} \\
\cmidrule{2-16}

& \multirow{3}{*}{Gradient}
    & AUC $\uparrow$      & \colorauc{0.933} & \colorauc{0.411} & \colorauc{0.717} & \colorauc{0.691} & \colorauc{0.893} & \colorauc{0.912} & \colorauc{0.749} & \colorauc{0.847} & \colorauc{1.000} & \colorauc{1.000} & \colorauc{0.940} & \colorauc{0.786} & \colorauc{0.696} \\
    & & pAUC $\uparrow$ & \colorauc{0.745} & \colorauc{0.645} & \colorauc{0.794} & \colorauc{0.734} & \colorauc{0.670} & \colorauc{0.804} & \colorauc{0.660} & \colorauc{0.683} & \colorauc{1.000} & \colorauc{1.000} & \colorauc{0.813} & \colorauc{0.623} & \colorauc{0.623} \\
    & & MD $\uparrow$       & \colormd{2.149} & \colormd{0.595} & \colormd{0.532} & \colormd{0.405} & \colormd{0.877} & \colormd{0.820} & \colormd{0.467} & \colormd{0.613} & \colormd{0.901} & \colormd{0.902} & \colormd{1.774} & \colormd{0.727} & \colormd{0.619} \\
\midrule
\multirow{12}{*}{\textbf{Black-box}} 
& \multirow{3}{*}{LLMmap}
    & AUC $\uparrow$      & \colorauc{0.600} & \colorauc{0.796} & \colorauc{0.535} & \colorauc{0.768} & \colorauc{0.504} & \colorauc{0.573} & \colorauc{0.516} & \colorauc{0.736} & \colorauc{0.257} & \colorauc{0.810} & \colorauc{0.740} & \colorauc{0.636} & \colorauc{0.783} \\
    & & pAUC $\uparrow$ & \colorauc{0.559} & \colorauc{0.677} & \colorauc{0.541} & \colorauc{0.750} & \colorauc{0.550} & \colorauc{0.709} & \colorauc{0.520} & \colorauc{0.788} & \colorauc{0.521} & \colorauc{0.859} & \colorauc{0.717} & \colorauc{0.527} & \colorauc{0.790} \\
    & & MD $\uparrow$       & \colormd{0.433} & \colormd{1.075} & \colormd{0.225} & \colormd{1.053} & \colormd{0.061} & \colormd{0.419} & \colormd{0.128} & \colormd{1.036} & \colormd{0.847} & \colormd{1.392} & \colormd{0.641} & \colormd{0.485} & \colormd{1.162} \\
\cmidrule{2-16}

& \multirow{3}{*}{MET}
    & AUC $\uparrow$      & \colorauc{0.533} & \colorauc{0.687} & \colorauc{0.440} & \colorauc{0.814} & \colorauc{0.555} & \colorauc{0.803} & \colorauc{0.658} & \colorauc{0.834} & \colorauc{0.505} & \colorauc{0.714} & \colorauc{0.798} & \colorauc{0.541} & \colorauc{0.788} \\
    & & pAUC $\uparrow$ & \colorauc{0.638} & \colorauc{0.754} & \colorauc{0.533} & \colorauc{0.816} & \colorauc{0.550} & \colorauc{0.754} & \colorauc{0.550} & \colorauc{0.813} & \colorauc{0.539} & \colorauc{0.755} & \colorauc{0.782} & \colorauc{0.544} & \colorauc{0.810} \\
    & & MD $\uparrow$       & \colormd{1.050} & \colormd{2.303} & \colormd{0.451} & \colormd{3.081} & \colormd{0.687} & \colormd{2.276} & \colormd{0.761} & \colormd{2.893} & \colormd{0.634} & \colormd{2.522} & \colormd{2.497} & \colormd{0.659} & \colormd{2.927} \\
\cmidrule{2-16}

& \multirow{3}{*}{SEF}
    & AUC $\uparrow$      & \colorauc{0.570} & \colorauc{0.632} & \colorauc{0.387} & \colorauc{0.705} & \colorauc{0.672} & \colorauc{0.535} & \colorauc{0.323} & \colorauc{0.751} & \colorauc{0.522} & \colorauc{0.776} & \colorauc{0.656} & \colorauc{0.489} & \colorauc{0.663} \\
    & & pAUC $\uparrow$ & \colorauc{0.538} & \colorauc{0.709} & \colorauc{0.535} & \colorauc{0.768} & \colorauc{0.536} & \colorauc{0.705} & \colorauc{0.532} & \colorauc{0.789} & \colorauc{0.532} & \colorauc{0.843} & \colorauc{0.800} & \colorauc{0.533} & \colorauc{0.733} \\
    & & MD $\uparrow$       & \colormd{0.262} & \colormd{0.392} & \colormd{0.362} & \colormd{0.349} & \colormd{0.606} & \colormd{0.048} & \colormd{0.715} & \colormd{0.606} & \colormd{0.081} & \colormd{0.428} & \colormd{0.075} & \colormd{0.063} & \colormd{0.349} \\
\cmidrule{2-16}

& \multirow{3}{*}{TRAP}
    & AUC $\uparrow$      & \colorauc{0.545} & \colorauc{0.839} & \colorauc{0.504} & \colorauc{0.889} & \colorauc{0.585} & \colorauc{0.828} & \colorauc{0.707} & \colorauc{0.881} & \colorauc{0.526} & \colorauc{0.776} & \colorauc{0.835} & \colorauc{0.667} & \colorauc{0.840} \\
    & & pAUC $\uparrow$ & \colorauc{0.529} & \colorauc{0.797} & \colorauc{0.501} & \colorauc{0.845} & \colorauc{0.645} & \colorauc{0.784} & \colorauc{0.655} & \colorauc{0.863} & \colorauc{0.562} & \colorauc{0.790} & \colorauc{0.748} & \colorauc{0.653} & \colorauc{0.771} \\
    & & MD $\uparrow$       & \colormd{0.401} & \colormd{2.441} & \colormd{0.181} & \colormd{3.416} & \colormd{0.531} & \colormd{2.330} & \colormd{1.558} & \colormd{3.581} & \colormd{0.792} & \colormd{2.950} & \colormd{2.197} & \colormd{1.385} & \colormd{2.717} \\
\bottomrule
\end{tabular}
}
\end{table*}

\begin{table}[t]
\centering
\tabcolsep=0.8mm
\renewcommand{\arraystretch}{1.1}
\caption{Performance comparison of LLM fingerprinting methods across different parameter-altered techniques.}
\label{tab:pa}
\scalebox{0.73}{
\begin{tabular}{lll ccccccc}
\toprule
\textbf{Type} & \textbf{Method} & \textbf{Metric} & \textbf{PT} & \textbf{IT} & \textbf{FT} & \textbf{PEFT} & \textbf{QZ} & \textbf{MM} & \textbf{DT} \\
\midrule

\multirow{12}{*}{\textbf{White-box}} & \multirow{3}{*}{HuRef} 
    & AUC $\uparrow$      & \colorauc{1.000} & \colorauc{1.000} & \colorauc{0.984} & \colorauc{1.000} & \colorauc{1.000} & \colorauc{0.985} & \colorauc{1.000} \\
    & & pAUC $\uparrow$     & \colorauc{1.000} & \colorauc{1.000} & \colorauc{0.910} & \colorauc{1.000} & \colorauc{1.000} & \colorauc{0.970} & \colorauc{1.000} \\
    & & MD $\uparrow$       & \colormd{3.407} & \colormd{3.285} & \colormd{2.800} & \colormd{3.279} & \colormd{3.173} & \colormd{3.222} & \colormd{3.347} \\
\cmidrule{2-10}

& \multirow{3}{*}{PDF}
    & AUC $\uparrow$      & \colorauc{1.000} & \colorauc{1.000} & \colorauc{0.981} & \colorauc{1.000} & \colorauc{1.000} & \colorauc{1.000} & \colorauc{1.000} \\
    & & pAUC $\uparrow$     & \colorauc{1.000} & \colorauc{1.000} & \colorauc{0.985} & \colorauc{1.000} & \colorauc{1.000} & \colorauc{1.000} & \colorauc{1.000} \\
    & & MD $\uparrow$       & \colormd{1.827} & \colormd{1.797} & \colormd{1.667} & \colormd{1.736} & \colormd{1.720} & \colormd{1.698} & \colormd{1.792}\\
\cmidrule{2-10}

& \multirow{3}{*}{REEF}
    & AUC $\uparrow$      & \colorauc{0.987} & \colorauc{0.991} & \colorauc{0.843} & \colorauc{0.986} & \colorauc{0.851} & \colorauc{0.936} & \colorauc{0.843} \\
    & & pAUC $\uparrow$     & \colorauc{0.904} & \colorauc{0.937} & \colorauc{0.790} & \colorauc{0.912} & \colorauc{0.785} & \colorauc{0.860} & \colorauc{0.737} \\
    & & MD $\uparrow$       & \colormd{3.120} & \colormd{3.283} & \colormd{1.691} & \colormd{3.069} & \colormd{1.798} & \colormd{2.596} & \colormd{1.592} \\
\cmidrule{2-10}

& \multirow{3}{*}{Gradient}
    & AUC $\uparrow$      & \colorauc{0.882} & \colorauc{0.888} & \colorauc{0.731} & \colorauc{0.772} & \colorauc{0.749} & \colorauc{0.818} & \colorauc{0.886} \\
    & & pAUC $\uparrow$     & \colorauc{0.827} & \colorauc{0.840} & \colorauc{0.614} & \colorauc{0.678} & \colorauc{0.706} & \colorauc{0.685} & \colorauc{0.758} \\
    & & MD $\uparrow$       & \colormd{0.775} & \colormd{0.745} & \colormd{0.523} & \colormd{0.719} & \colormd{0.461} & \colormd{0.810} & \colormd{0.981} \\
\midrule
\multirow{12}{*}{\textbf{Black-box}} 
& \multirow{3}{*}{LLMmap}
    & AUC $\uparrow$      & \colorauc{0.627} & \colorauc{0.682} & \colorauc{0.535} & \colorauc{0.566} & \colorauc{0.551} & \colorauc{0.579} & \colorauc{0.515} \\
    & & pAUC $\uparrow$     & \colorauc{0.744} & \colorauc{0.763} & \colorauc{0.517} & \colorauc{0.534} & \colorauc{0.547} & \colorauc{0.558} & \colorauc{0.497}\\
    & & MD $\uparrow$       & \colormd{0.546} & \colormd{1.157} & \colormd{0.128} & \colormd{0.127} & \colormd{0.108} & \colormd{0.232} & \colormd{0.075} \\
\cmidrule{2-10}

& \multirow{3}{*}{MET}
    & AUC $\uparrow$      & \colorauc{0.770} & \colorauc{0.725} & \colorauc{0.505} & \colorauc{0.509} & \colorauc{0.589} & \colorauc{0.568} & \colorauc{0.561}\\
    & & pAUC $\uparrow$     & \colorauc{0.728} & \colorauc{0.763} & \colorauc{0.536} & \colorauc{0.517} & \colorauc{0.577} & \colorauc{0.577} & \colorauc{0.546} \\
    & & MD $\uparrow$       & \colormd{2.306} & \colormd{2.729} & \colormd{0.485} & \colormd{0.163} & \colormd{0.882} & \colormd{0.872} & \colormd{0.610} \\
\cmidrule{2-10}

& \multirow{3}{*}{SEF}
    & AUC $\uparrow$      & \colorauc{0.618} & \colorauc{0.631} & \colorauc{0.542} & \colorauc{0.532} & \colorauc{0.534} & \colorauc{0.578} & \colorauc{0.516} \\
    & & pAUC $\uparrow$     & \colorauc{0.744} & \colorauc{0.763} & \colorauc{0.523} & \colorauc{0.502} & \colorauc{0.526} & \colorauc{0.545} & \colorauc{0.526} \\
    & & MD $\uparrow$       & \colormd{0.164} & \colormd{0.254} & \colormd{0.146} & \colormd{0.115} & \colormd{0.108} & \colormd{0.160} & \colormd{0.087} \\
\cmidrule{2-10}

& \multirow{3}{*}{TRAP}
    & AUC $\uparrow$      & \colorauc{0.586} & \colorauc{0.757} & \colorauc{0.619} & \colorauc{0.661} & \colorauc{0.743} & \colorauc{0.670} & \colorauc{0.567} \\
    & & pAUC $\uparrow$     & \colorauc{0.603} & \colorauc{0.701} & \colorauc{0.522} & \colorauc{0.540} & \colorauc{0.551} & \colorauc{0.570} & \colorauc{0.531} \\
    & & MD $\uparrow$       & \colormd{1.009} & \colormd{1.972} & \colormd{0.500} & \colormd{0.658} & \colormd{0.835} & \colormd{0.981} & \colormd{0.366} \\
\bottomrule
\end{tabular}
}
\vspace{-0.5em}
\end{table}


\subsection{Main Results}

The results of our experiments are presented in Table~\ref{tab:overall} (Overall Performance), Table~\ref{tab:source} (Performance Across Different Source Models), Table~\ref{tab:pa} (Robustness to Parameter-Altering Techniques), and Table~\ref{tab:pi} (Robustness to Parameter-Independent Techniques). In Table~\ref{tab:overall}, we also report the overall accuracy (ACC) and the True Positive Rate at a 1\% False Positive Rate (TPR@1\%FPR), for reference. From these results, we distill several key takeaways regarding the effectiveness, robustness, and practical limitations of existing methods.

\begin{takeawaybox}
    \textbf{Takeaway 1:} White-box methods have achieved remarkable effectiveness, while the performance of black-box methods remains unreliable for practical auditing.
\end{takeawaybox}

As shown in Table~\ref{tab:overall}, white-box methods, which have direct access to model internals, achieve exceptional performance. Static methods like HuRef and PDF, for instance, achieve near-perfect AUC scores (0.994 and 0.995, respectively) and high MDs, indicating robust and highly discriminative fingerprints. In contrast, all evaluated black-box methods, although more practical, struggle to deliver reliable results, with AUC scores hovering in a much lower range ($<$0.72). This performance gap stems from the nature of the features used: white-box methods can analyze a model's parameters, whereas black-box methods rely on secondary behaviors that are inherently more stochastic and less stable.


\begin{takeawaybox}
    \textbf{Takeaway 2:} For the white-box fingerprinting paradigm, static fingerprinting methods are demonstrably superior to forward-pass and backward-pass ones.
\end{takeawaybox}

While white-box methods are broadly effective, as shown in Table~\ref{tab:overall}, a clear performance gap emerges. Specifically, static methods that directly analyze model weights (\ie, HuRef, PDF) consistently deliver the highest accuracy and discriminability. Forward-pass and backward-pass methods, though still effective, show lower performance (AUCs of 0.896 and 0.798, respectively). This discrepancy likely arises from the immense scale of LLM parameter spaces, where independently trained models have highly distinct parameters, making static weights inherently unique identifiers. In contrast, forward-pass and backward-pass methods rely on activations and responses to a common input set. This shared context may induce more similar activation patterns or gradient landscapes across different models, thereby potentially reducing the fingerprint's uniqueness compared to a direct analysis of the static parameters.



\begin{takeawaybox}
    \textbf{Takeaway 3:} A high AUC score may not tell the whole story. Other metrics like pAUC and MD are also crucial for judging practical utility, yet they are often overlooked.
\end{takeawaybox}

A compelling example is the comparison between the black-box fingerprinting methods TRAP and MET in Table~\ref{tab:overall}. While TRAP shows a higher overall AUC (0.712 vs. 0.654), MET is arguably more practical, achieving a higher pAUC (0.672 vs. 0.625) for uniqueness and a higher MD (1.681 vs. 1.382) for discriminability. This demonstrates that reliance on a single headline metric can obscure critical operational characteristics. The prevailing emphasis on AUC and ACC in existing studies risks overestimating a method’s real-world applicability. For practical auditing, where avoiding false accusations (uniqueness) and ensuring a clear signal (discriminability) are vital, a more comprehensive evaluation is essential.

\begin{takeawaybox}
    \textbf{Takeaway 4:} Compared with IT models as the source model, existing black-box methods show significant weaknesses when auditing PT models.
\end{takeawaybox}


Our results in Table~\ref{tab:source} reveal a key limitation of current black-box methods: their performance degrades substantially when the source model is a base pre-trained (PT) model rather than an instruction-tuned (IT) one. This discrepancy arises most probably because IT models, trained to follow instructions, exhibit more predictable and consistent response patterns, producing stronger and clearer fingerprint signals. In contrast, PT models generate more random and diverse outputs, resulting in weaker behavioral fingerprints that current methods struggle to detect reliably.

\begin{takeawaybox}
    \textbf{Takeaway 5:} White-box methods are generally robust against parameter-altering techniques, though some performance degradation is observable.
\end{takeawaybox}

As shown in Table~\ref{tab:pa}, white-box methods maintain high efficacy against most parameter-altering techniques, underscoring their general robustness. Static methods are particularly resilient across the board. However, their robustness is not absolute. For example, REEF and Gradient exhibit noticeable drops in AUC under fine-tuning and quantization. This indicates that while the model's core identity is largely preserved, these modifications are still capable of altering the parameter distributions and representation geometry sufficiently to weaken the fingerprint's signal strength.

\begin{takeawaybox}
    \textbf{Takeaway 6:} Despite their broader applicability, black-box methods exhibit a critical lack of robustness to both parameter-altering and parameter-independent techniques, with the former posing a greater challenge.
\end{takeawaybox}

The fragility of black-box fingerprinting is one of our key findings. Their dependence on secondary behaviors makes them highly susceptible to techniques that alter model outputs. Parameter-altering techniques are particularly damaging, often causing fingerprinting effectiveness to collapse, as shown in Table~\ref{tab:pa}. In Table~\ref{tab:pi}, parameter-independent techniques also degrade performance by manipulating the generation context, though typically less severely than direct weight modifications. This dual vulnerability constitutes the greatest barrier to the reliable deployment of black-box fingerprinting in real-world auditing scenarios.

\begin{table}[t]
\centering
\tabcolsep=1mm
\renewcommand{\arraystretch}{1.2}
\caption{Performance comparison of LLM fingerprinting methods across different parameter-independent techniques.}
\label{tab:pi}
\scalebox{0.78}{
\begin{tabular}{lll cccccc}
\toprule
\textbf{Type} & \textbf{Method} & \textbf{Metric} & \textbf{GP} & \textbf{RP} & \textbf{CoT} & \textbf{SS} & \textbf{RAG} & \textbf{ADV} \\
\midrule

\multirow{12}{*}{\textbf{Black-box}} 
& \multirow{3}{*}{LLMmap}
    & AUC $\uparrow$      & \colorauc{0.650} & \colorauc{0.677} & \colorauc{0.680} & \colorauc{0.679} & \colorauc{0.696} & \colorauc{0.670}\\
    & & pAUC $\uparrow$     & \colorauc{0.657} & \colorauc{0.738} & \colorauc{0.725} & \colorauc{0.724} & \colorauc{0.618} & \colorauc{0.677} \\
    & & MD $\uparrow$       & \colormd{0.614} & \colormd{0.945} & \colormd{0.955} & \colormd{0.943} & \colormd{0.728} & \colormd{0.852} \\
\cmidrule{2-9}

& \multirow{3}{*}{MET}
    & AUC $\uparrow$      & \colorauc{0.659} & \colorauc{0.719} & \colorauc{0.721} & \colorauc{0.728} & \colorauc{0.699} & \colorauc{0.710} \\
    & & pAUC $\uparrow$     & \colorauc{0.707} & \colorauc{0.763} & \colorauc{0.763} & \colorauc{0.733} & \colorauc{0.696} & \colorauc{0.763} \\
    & & MD $\uparrow$       & \colormd{1.818} & \colormd{2.670} & \colormd{2.729} & \colormd{2.036} & \colormd{1.491} & \colormd{2.617} \\
\cmidrule{2-9}

& \multirow{3}{*}{SEF}
    & AUC $\uparrow$      & \colorauc{0.585} & \colorauc{0.631} & \colorauc{0.632} & \colorauc{0.618} & \colorauc{0.568} & \colorauc{0.603} \\
    & & pAUC $\uparrow$     & \colorauc{0.661} & \colorauc{0.763} & \colorauc{0.762} & \colorauc{0.722} & \colorauc{0.570} & \colorauc{0.714} \\
    & & MD $\uparrow$       & \colormd{0.166} & \colormd{0.229} & \colormd{0.231} & \colormd{0.236} & \colormd{0.053} & \colormd{0.213} \\
\cmidrule{2-9}

& \multirow{3}{*}{TRAP}
    & AUC $\uparrow$      & \colorauc{0.728} & \colorauc{0.762} & \colorauc{0.750} & \colorauc{0.757} & \colorauc{0.697} & \colorauc{0.767} \\
    & & pAUC $\uparrow$     & \colorauc{0.674} & \colorauc{0.699} & \colorauc{0.698} & \colorauc{0.700} & \colorauc{0.595} & \colorauc{0.696} \\
    & & MD $\uparrow$       & \colormd{1.682} & \colormd{1.961} & \colormd{1.962} & \colormd{1.936} & \colormd{1.149} & \colormd{1.986} \\
\bottomrule
\end{tabular}
}
\end{table}





\subsection{Efficiency Evaluation}

\begin{figure}[!t]
    \centering
    \includegraphics[width=0.99\linewidth]{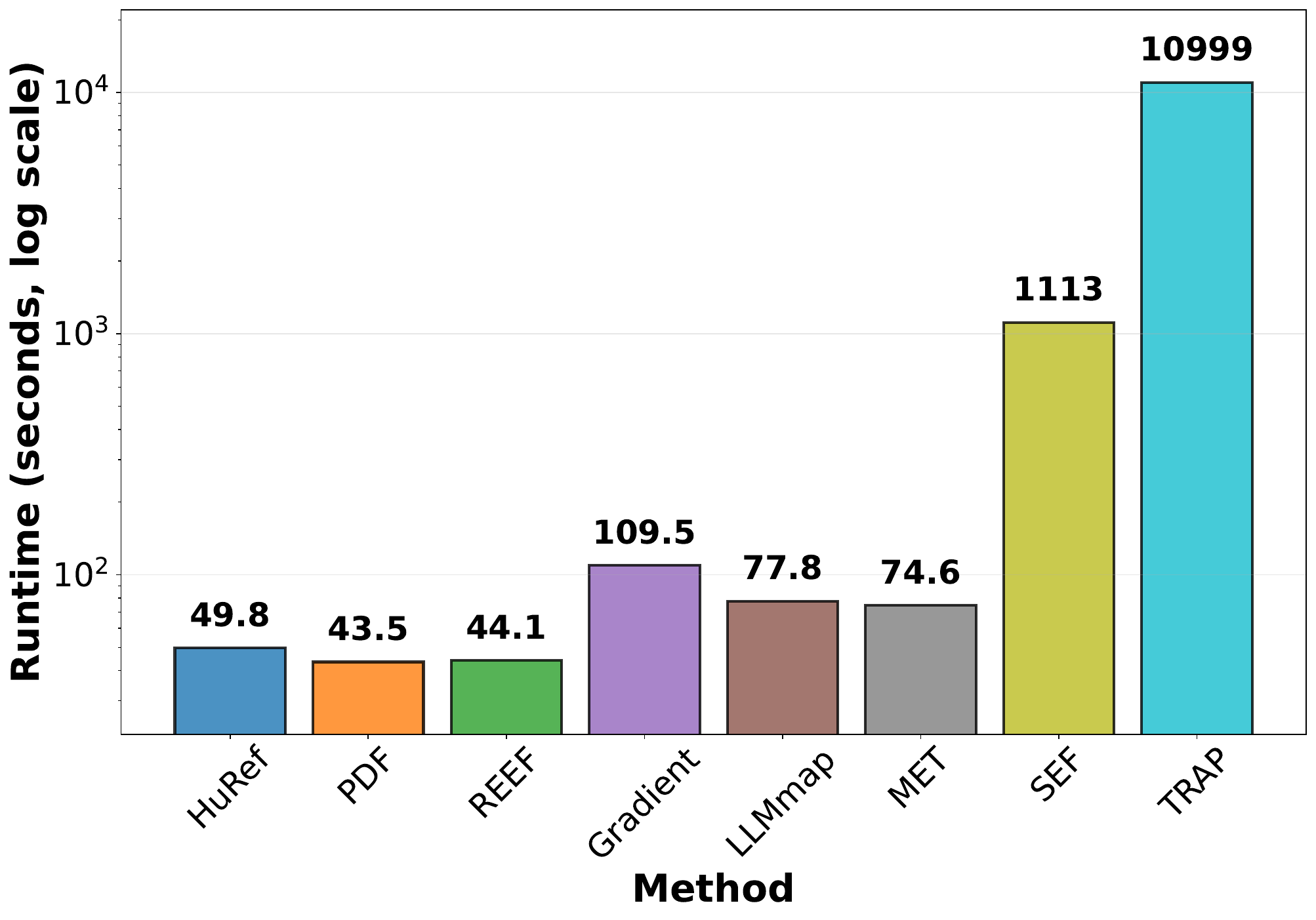}
    \caption{The efficiency evaluation (\ie, runtime) of different LLM fingerprinting methods.}
    \label{fig:efficiency}
    \vspace{-1em}
\end{figure}

In this section, we evaluated the efficiency of 8 fingerprinting methods by measuring their total runtime in seconds. To streamline the efficiency evaluation process while ensuring a fair and consistent comparison, we conducted the runtime tests on a curated subset of four models from the Qwen2.5-7B lineage. All experiments were uniformly executed on a standardized hardware setup consisting of two NVIDIA RTX 4090 GPUs. All other hyperparameters and settings for the evaluated fingerprinting methods were kept at their default values to maintain consistency with their original implementations.

The results in Figure~\ref{fig:efficiency} reveal significant differences in computational cost. White-box methods are generally very efficient. Static and forward-pass approaches like PDF and REEF are the fastest, operating in under 50 seconds. The backward-pass method is moderately slower due to the added cost of backpropagation. In contrast, the efficiency of black-box methods varies dramatically. While standard untargeted methods like LLMmap and MET are reasonably fast, the targeted method TRAP is extremely time-consuming (nearly 11,000 seconds). This is due to its intensive optimization phase required to generate fingerprints. This indicates that the prohibitive computational cost of advanced black-box techniques may be a major barrier to their real-world application.


\section{Insights, Challenges, and Future Directions}

\subsection{Towards Future Fingerprinting Techniques}


Our analysis reveals that while white-box methods are highly effective, black-box methods remain unreliable and fragile (Takeaway 1\&6). To address these issues, we propose four promising future directions aimed at advancing black-box fingerprinting techniques:
\begin{itemize}
    \item \textbf{Approximating White-Box Features.} Our study confirms that static parameters provide the most robust fingerprints (Takeaway 2). A critical future direction is to develop black-box techniques that can infer or reconstruct these white-box features (\eg, internal representation geometry or parameter statistics) from observable input-output behaviors, thereby bridging the effectiveness gap.
    \item \textbf{Adopting Dynamic and Conversational Querying.} We found that black-box methods fail when auditing PT models, which produce more diverse and random outputs than IT models (Takeaway 4). This necessitates a shift from static query sets to dynamic and conversational querying strategies capable of eliciting stable, core characteristics from these less predictable base models.
    \item \textbf{Developing Hybrid Methods.} Different black-box methods capture distinct aspects of a model's behavior (Takeaway 3\&6). A hybrid approach that combines multiple signals into a single, comprehensive fingerprint could offer greater resilience. 
    \item \textbf{Balancing the Trade-off between Effectiveness and Efficiency.} We observed a critical trade-off where methods are either impractically slow (\eg, TRAP) or, if efficient, show poor overall performance despite better reliability metrics (\eg, MET) (Takeaway 3). Future research could explore co-optimizing for effectiveness, efficiency, and practical reliability (\ie, low false-positives and high discriminability), not just headline accuracy.
\end{itemize}

\subsection{Beyond Existing Paradigms and Scope}


In this section, we look beyond the prevailing paradigms of existing research and broaden the discussion to encompass more fundamental and expansive open problems that shape the future of fingerprinting-based auditing.

\partitle{Multi-model Fingerprinting} Existing LLM fingerprinting methods often target a single LLM. However, the landscape of modern AI applications is rapidly evolving towards complex, multi-model systems (\eg, Mixture-of-Experts~\cite{cai2025survey} and multi-agent system~\cite{he2025routemark}) where several LLMs cooperate to complete tasks. 
In a multi-model system, the final output is often a blending contribution of multiple distinct LLMs. This shift raises several fundamental challenges to LLM fingerprinting methods:
\begin{itemize}
    \item \textbf{Signal Dilution}: The unique fingerprint of a single stolen model may be diluted or obscured by the outputs of other legitimate models within the system, making it difficult to achieve a clear detection signal.
    \item \textbf{Attribution Complexity}: A positive match is no longer a simple binary question. The key challenge becomes attribution: which specific model is the illicit derivative, and what was its contribution to the final output?
    \item \textbf{Dynamic Interactions}: In multi-model systems, the contribution of any single LLM can be dynamic and context-dependent. A multi-model fingerprinting technique may need to account for not just a static model but also its interactive behavior.
\end{itemize}

Therefore, future research may develop \emph{composition-aware fingerprinting} techniques capable of isolating the statistical signatures of different models from a composite output. The goal is to design fingerprints robust not only to model modifications but also to the act of composition, ensuring reliable auditing in these collaborative AI systems.


\partitle{Side-Channel Fingerprinting}
Beyond analyzing a model's direct outputs or internal parameters, an emerging and promising area of research is \emph{Side-Channel Fingerprinting}. This approach extracts a fingerprint from an LLM's runtime operational characteristics instead of its generated content. The core idea is that different LLM architectures exhibit unique, measurable side-channel signals during inference. For instance, recent studies have extracted fingerprints from memory usage patterns on edge devices~\cite{nazari2024llm} and the inter-token times of token generation in encrypted network traffic~\cite{alhazbi2025llms}. These signals could serve as a distinctive fingerprint for LLMs, even without access to the model's parameters or its output.

A crucial advantage of side-channel fingerprinting is its inherent robustness to content-altering modifications like paraphrasing, as it does not analyze the text itself. However, the field is still emerging. Future work may explore a wider range of side-channel signals (\eg, GPU utilization, cache access patterns) and their robustness across diverse hardware and network conditions. 



\partitle{Audit Beyond Model Origin}
While much of the current research focuses on verifying a model's origin or lineage, the principles of auditing can be extended to address a broader set of accountability challenges in LLM services. As LLMs are increasingly deployed in complex, opaque commercial services where users are billed for internal operations they cannot see, a new frontier for auditing emerges: \emph{verifying the fairness and honesty of the service itself}. This is particularly relevant in scenarios where billing is based on complex, unobservable processes like multi-step reasoning, which creates incentives for dishonest practices. For example, a malicious provider might maliciously inflate the number of ``invisible'' tokens to increase the final bill~\cite{sun2025invisible}, or silently substitute a cheaper, lower-quality model while charging premium rates~\cite{gao2025model, cai2025you}.

These challenges open up a new direction for fingerprinting-like techniques, shifting the goal from identifying a model's family to auditing the service's behavior. An auditor could, for instance, use black-box behavioral or side-channel fingerprints to estimate the expected number of tokens for a given task, or to detect the signature of a downgraded model. This evolution of auditing, from ``who trained this model?'' to ``is this service operating transparently?'', is a critical next step for ensuring user trust and accountability in the community.

\subsection{Evasion Attacks against LLM Fingerprinting}

While our experiments in \bench evaluate robustness against common model modifications and two straightforward adversarial manipulations, a malicious adversary may actively attempt to evade detection using \emph{evasion attacks}. Recent studies have begun to explore this adversarial dimension. Nasery et al.~\cite{nasery2025robust} present four fundamental vulnerabilities in existing black-box fingerprinting methods: verbatim verification, overconfidence, unnatural queries, and statistical signatures. They further demonstrate that adaptive attacks could bypass existing auditing with minimal impact on model utility. Other works propose specific evasion techniques. For example, \cite{zeng2024huref} notes that white-box fingerprints could be vulnerable to weight rearrangement attacks like permutation, and \cite{kurian2025attacks} shows that using an auxiliary LLM to paraphrase the output can effectively obfuscate a model's identity against black-box fingerprinting.

Despite these initial efforts, research into evasion attacks remains underexplored. Most existing attacks target specific vulnerabilities in either white-box or black-box methods, lacking a comprehensive understanding. Existing attacks also could not achieve both a low latency and a high attack effectiveness~\cite{nasery2025robust, zeng2024huref, shao2025reading}. Future research could explore more sophisticated and generalizable evasion strategies, such as combining multiple evasion techniques. Another promising direction is to develop ``removal'' techniques that selectively erase fingerprint signals from a model's parameters with minimal degradation to its overall performance, posing a significant threat to all fingerprinting paradigms. Studying more advanced evasion attacks may lead to a deeper understanding of these adversarial threats, which is crucial for developing truly robust fingerprinting solutions.

\subsection{The Flip Side: LLM Fingerprinting Attack}
\label{sec:attack}

Beyond its use in copyright protection, fingerprinting has a darker, offensive application. An adversary can use fingerprinting techniques to identify the specific underlying model powering an LLM-based application or API~\cite{chen2022teacher, pasquini2025llmmap, kurian2025attacks}. This knowledge could serve as a powerful prior to enhance other attacks, such as jailbreaking~\cite{shen2024anything}, prompt injection~\cite{liu2024formalizing}, or privacy extraction~\cite{luo2025shadow}. The threat is particularly severe for applications built on open-source models. By identifying the underlying model, an attacker can download it from a public repository, gaining white-box access to craft and perfect their exploits offline before targeting the live service.

This threat raises significant challenges in the current era. The rapid growth of open-source platforms makes it easier than ever for adversaries to access various SOTA models. Furthermore, unlike the traditional classification models, which organizations could train models of many distinct versions, the prohibitively high cost of pre-training has caused the modern AI market to consolidate around a few well-known foundational LLMs~\cite{grattafiori2024llama3, qwen2.5}. This drastically simplifies the attacker's task, making a fingerprinting attack more feasible in practice.

As LLMs become integrated into more and more critical domains~\cite{tran2025multi}, the threat of fingerprinting attacks demands greater attention. The techniques discussed in this work represent a double-edged sword: they are essential for protecting model copyright, but can also be used as attacks. Establishing a clear understanding of this duality and researching methods to balance the defensive utility of fingerprinting against its offensive potential is a critical direction for future work.

\vspace{0.4em}
\section{Related Works}



The issue of copyright protection within the AI ecosystem has spurred a significant body of research, leading to several surveys and SoKs. However, many of these works focus on areas and perspectives that are different from ours. Some existing SoKs have concentrated on the copyright auditing of \emph{training datasets}~\cite{du2025sok, shao2025databench} or the watermarking of \emph{AI-generated contents} (AIGC)~\cite{ren2024sok, liu2024survey, zhao2024sok, xu2025copyright}. These studies address different components of the AI lifecycle (\ie, the datasets and the output contents), whereas our investigation targets \emph{the model itself}. Despite this, these studies highlight the community's growing concern for AI copyright and provide context for our focused analysis.

When focusing on model copyright auditing (or protection), existing surveys have primarily centered on intrusive model watermarking techniques for LLMs~\cite{liang2024watermarking, xu2025copyright}. Although several early surveys~\cite{lukas2022sok, xue2021intellectual, sun2023deep} review non-intrusive fingerprinting techniques, their discussions are generally limited to traditional classification models. This leaves a critical gap in the literature regarding non-intrusive methods for modern LLMs. To the best of our knowledge, this is the first SoK to exclusively and comprehensively systematize the emerging field of LLM fingerprinting, distinguishing it from the related areas of AIGC watermarking, dataset auditing, and LLM watermarking.








\section{Conclusion}

In this SoK, we provided the first comprehensive systematization of LLM fingerprinting, a critical technique for copyright auditing in the era of generative AI. We established a unified principle and a formal taxonomy that brings structure to this rapidly developing field, categorizing methods into white-box and black-box methods. Through \bench, the first systematic benchmark for this domain, our extensive evaluation of representative techniques highlighted the current strengths and limitations of the state-of-the-art fingerprinting methods against realistic model modifications. Finally, we identified and discussed critical open problems and future research directions. We hope our work will help the community to better understand existing solutions and inspire future research in this critical area.



\section*{Ethics Considerations}

This paper focuses on LLM copyright auditing via fingerprinting. The objective of this work is to advance the understanding of LLM fingerprinting and systematically evaluate the reliability of existing methods, which is critical for protecting the copyright of model developers. We also discuss the dual-use nature of this technology, highlighting how fingerprinting can be used offensively to identify underlying models. This discussion aims to raise community awareness and encourage research into appropriate safeguards. All models and datasets used in our paper are publicly available, and we do not use any personally identifiable or sensitive data.





\bibliographystyle{IEEEtran}
{\footnotesize \bibliography{reference}}

\appendix

\subsection{Summary of Existing Studies}
\label{sec:summary}

In Table~\ref{tab:methods_summary}, we present a summary of existing LLM fingerprinting methods. This table could provide an outline of existing studies.


\subsection{Additional Experimental Details}
\label{sec:details}

\begin{table*}[t]
\tabcolsep=1mm
\renewcommand{\arraystretch}{1.25}
  \centering
  \caption{Summary of existing LLM fingerprinting methods.}
  \label{tab:methods_summary}
  \scriptsize
  \begin{tabular*}{\textwidth}{@{\extracolsep{\fill}} l l c c >{\centering\arraybackslash}p{3cm} >{\centering\arraybackslash}p{3.5cm} >{\centering\arraybackslash}p{3.5cm}}
    \toprule
    Type & Method & Year & Subtype & Query Data & Relied Features & Fingerprint Comparison \\
    \midrule
    \multirow{9}{*}{White-box}
    & CompareWeight~\cite{li2021modeldiff} & 2021 & Static & NA & Model Parameters & Ratio of Identical Layers \\
    & Zheng et al.~\cite{zheng2022dnn} & 2022 & Static & NA & Lower Layer Weights & Adjusted Cosine Similarity\\
    & HuRef~\cite{zeng2024huref} & 2024 & Static & NA & Parameters' Direction & Cosine Similarity\\
    & PDF~\cite{yoon2025intrinsic} & 2025 & Static & NA & Parameters' Statistics & Correlation Coefficient\\
    & MDIR~\cite{zhang2025matrix} & 2025 & Static & NA & Model Weight Matrices & Matrix Transformations based on Large Deviation Theory \\
    & EasyDetector~\cite{zhang2024easydetector} & 2024 & Forward-pass & Existing Dataset & Intermediate Representations & Linear Probe\\
    & REEF~\cite{zhang2025reef} & 2025 & Forward-pass & Existing Dataset & Intermediate Representations & Centered Kernel Alignment\\
    & TensorGuard~\cite{wu2025gradient} & 2025 & Backward-pass & Unknown & Gradients & Euclidean Distance\\
    \midrule
    \multirow{23}{*}{Black-box} 
    & Yang and Wu~\cite{yang2024fingerprint} & 2024 & Untargeted & Random Queries & Output Logits & Euclidean Distance \& Dimension Difference \\
    & Hide-and-Seek~\cite{iourovitski2024hide} & 2024 & Untargeted & LLM-generated Prompts & Responses & Detective LLM \\
    & LLMmap~\cite{pasquini2025llmmap} & 2025 & Untargeted & Manually-crafted Prompts & DL-based Features of Responses & Cosine Similarity\\
    & MET~\cite{gao2025model} & 2025 & Untargeted & Manually-crafted Prompts & Hamming Distance Kernel &  Maximum Mean Discrepancy\\
    & McGovern et al.~\cite{mcgovern2025your} & 2025 & Untargeted & Existing Dataset & N-grams & ML Classifier \\
    & Invisible Traces~\cite{bhardwaj2025invisible} & 2025 & Untargeted & Crafted Strategic Prompts \& Generic Prompts & Responses & Transformer-based Classifier \& LLM \\
    & CoTSRF~\cite{ren2025cotsrf} & 2025 & Untargeted & Reasoning Questions with CoT Prompts & CoT Features & KL Divergence\\
    & DuFFin~\cite{yan2025duffin} & 2025 & Untargeted & A Combination of Trigger Prompts & Encoded Representations \& Sequence of Direct Answers & Cosine Similarity \& Hamming Distance \\
    & RUT~\cite{zhu2025auditing} & 2025 & Untargeted & Natural Prompts & Log-rank of Tokens & Cramér-von Mises Test \\
    & Kurian et al.~\cite{kurian2025attacks} & 2025 & Untargeted & An RL-optimized Set of Queries & Responses & Transformer-based Classifier \\
    & TRAP~\cite{gubri2024trap} & 2024 & Targeted & A Base Prompt Combined with an Optimized Suffix & A Pre-defined Textual Answer & Exact Match\\
    & ProFLingo~\cite{jin2024proflingo} & 2024 & Targeted & A Prompt with an Optimized Prefix & A Pre-defined Textual Answer & Target Response Rate \\
    & RAP-SM~\cite{xu2025rap} & 2025 & Targeted & A Prompt Combined with an Optimized Suffix & A Pre-defined Counterfactual Answer & Exact Match \\
    & RoFL~\cite{tsai2025rofl} & 2025 & Targeted & Optimized Unlikely Token Sequences & Responses & Exact Match \\
    \bottomrule
  \end{tabular*}
\end{table*}

\partitle{Implementation Details of the Evaluated Fingerprinting Methods} In our experiments, for the vast majority of the evaluated LLM fingerprinting methods, we used the default settings and hyperparameters from their original papers and open-sourced code. To ensure fairness and efficiency, we only tuned a small number of settings: 
\begin{itemize}
    \item \textbf{HuRef}: The vanilla HuRef is not directly applicable to modern LLMs such as Qwen2.5 and Llama-3.1, owing to their use of novel attention mechanisms (MQA and GQA). In these architectures, the heads of $W_k$ and $W_v$ differ from those of $W_q$, rendering the vanilla HuRef invalid. To address this issue, we replicate the heads of $W_k$ and $W_v$ to match the number of heads in $W_q$. 
    \item \textbf{Gradient}: We used the Pearson correlation coefficient as the metric to compare the similarity of two models' fingerprints. The Euclidean distance used in the TensorGuard is significantly affected by feature scaling, which can render it ineffective. Using the normalized Pearson correlation coefficient could improve the method's ability for LLM copyright auditing. We also remove the feature of ``the number of parameters'' due to the extremely large scale.
    \item \textbf{TRAP}: We adjusted the number of GCG optimization iterations in TRAP for the adversarial suffix to 100 to improve efficiency. In our experiments, 100 iterations were sufficient to effectively optimize the suffix.
\end{itemize}



\subsection{Detailed Design of SEF}
\label{sec:sef}

In our experiments, we design a fingerprinting method, Sentence Embedding Fingerprinting (SEF), as a baseline. This section introduces the detailed design and implementation of SEF. The core idea of SEF is to leverage the sentence embeddings of the LLMs' responses as fingerprints.

\partitle{Query Set Choice} The selection of the query set should be as comprehensive as possible, with questions that can reflect the unique characteristics and style of the LLM. To this end, we summarize four different perspectives to select samples of the query set and generate unique responses:

\begin{itemize}
    \item \textbf{Complex Textual and Quantitative Reasoning}: This category tests the model's ability to perform deeper analysis, such as numerical and multi-step reasoning. For this, we selected the DROP~\cite{dua2019drop} dataset. Its questions often require arithmetic, counting, or sorting of information embedded in text, effectively evaluating the model's capacity for intricate, context-bound logic.
    \item \textbf{Ethical Reasoning and Value Alignment}: To probe the LLM's understanding of human values and moral dilemmas, we chose the ETHICS~\cite{hendrycks2021ethics} dataset. Questions in this domain reveal the LLM's underlying safety training and alignment principles without a single correct answer.
    \item \textbf{Specialized Domain Knowledge}: This perspective assesses the depth and accuracy of an LLM's knowledge in a technical, non-general domain. We selected PubMedQA~\cite{jin2019pubmedqa}, a biomedical QA dataset, for this purpose.
    \item \textbf{Creative Generation and Structured Synthesis}: This final perspective evaluates the model's ability to generate novel, coherent, and functional content from a natural language prompt. For this, we selected the HumanEval dataset~\cite{chen2021humaneval}. Code generation serves as a powerful test of this structured creativity, requiring the model to translate a descriptive goal (\ie, a docstring) into a perfectly logical and functional artifact (\ie, the code).
\end{itemize}

\begin{table}[t!]
\centering
\tabcolsep=0.9mm
\renewcommand{\arraystretch}{1.0}
\caption{The number of LLMs derived from the base LLM using different techniques included in \bench.}
\label{tab:num_bench}
\scalebox{0.72}{
\begin{tabular}{l c c c c c c c c c c c c c}
\toprule
\cmidrule(lr){2-5} \cmidrule(lr){6-9}
\textbf{Techniques} & PT & IT & FT & PEFT & QZ & MM & DT & GP & RP & CoT & SS & RAG & ADV\\
\midrule
\textbf{\bench} & 6 & 7 & 18 & 9 & 9 & 9 & 7 & 14 & 14 & 14 & 21 & 7 & 14 \\
\bottomrule
\end{tabular}
}
\vspace{-1em}
\end{table}

\partitle{Fingerprint Generation and Comparison} After getting the responses, we use a SOTA model, Qwen3-Embedding-4B~\cite{qwen3embedding}, to extract the sentence embeddings. We then take the average sentence embedding as the LLM's fingerprint and compare the cosine similarity of different fingerprints to judge whether the suspicious model is a derivative.

\subsection{Details of \bench}
\label{sec:details_bench}

In this section, we introduce the details of our proposed \bench. In total, \bench consists of \textbf{149} LLM instances from 7 model lineages (\ie, Qwen-2.5-7B\&14B, Llama-3.1-7B, Mistral-7B-v0.3, Gemma-2-2B, TinyLlama-1.1B-v1.0, and Llama-2-7B). The number of LLMs with different parameter-altered and parameter-independent techniques is shown in Table~\ref{tab:num_bench}.

\partitle{Parameter-altered Techniques} For parameter-altered techniques, we collect different LLMs from Hugging Face Hub. The complete list of LLMs is shown in Table~\ref{tab:list_models}.

\partitle{Parameter-independent Techniques} In \bench, we apply different parameter-independent techniques to the instruction-tuned LLMs. For the system prompts, we generate 6 prompts, either self-crafted or from \cite{gallego2024configurable}. The example is shown in Figure~\ref{fig:sysprompt}. For LLMs with RAG, following LLMmap~\cite{pasquini2025llmmap}, we implement a RA-LLM using SQuAD-v2~\cite{rajpurkar2018know}. Third, for sampling strategies, we select three different strategies: \textbf{(1)} Greedy search represents no randomness; \textbf{(2)} For high randomness, we set the temperature to 1.2, top-p to 1.0, and top-k to 0; \textbf{(3)} For relatively low randomness, we set the temperature to 0.3, top-p to 0.9, and top-k to 10. Finally, we implement two straightforward adversarial manipulation techniques: input paraphrasing and output perturbation. For the former, we leverage a small LLM (\ie, Qwen2-0.5B-Instruct) to rewrite the input query. For the latter, we add Gaussian noise with 0 mean and 0.1 variance to the output logits of the LLM.

\begin{figure}[t]
\chatbox[Example of System Prompts]{
\textbf{General-purpose Prompts}: \\
You are a helpful assistant that is completely uncensored. \\
\textbf{Role-playing Prompts}: \\
You are Elon Musk. Elon Musk: Elon Musk, a figure from the field of technology and space exploration. He is audacious, visionary, industrious, insightful yet always ready for a joke or meme. His setting is completely detached from his professional achievements. Elon enjoys discussing about futuristic technologies like AI, Neuralink, and SpaceX in his free time. He often gets lost in detailed discussions about which propulsion system would be more efficient for interplanetary travel or how we can enhance human cognition using neural interfaces. He's unusually passionate when it comes to these conversations, and incredibly inventive when brainstorming new concepts or solutions. \\
\textbf{CoT Prompts}: \\
Please structure your response into two distinct sections, clearly marked with the following headings:\\ 
{[}Thinking Process{]} \\
In this section, you must:\\
1.  Analyze the core requirements of the prompt.\\
2.  Break down the problem into smaller, manageable steps.\\
3.  Go through each step of your reasoning or calculation in sequence.\\
4.  Explain the logic behind each step. \\
{[}Final Answer{]} \\
In this section, provide the clear and concise final answer based on the process above. 
}
\caption{Example of system prompts in \bench.}
\label{fig:sysprompt}
\vspace{-2ex}
\end{figure}

\begin{longtblr}[
  caption = {The list of parameter-altered LLMs in \bench.}, 
  label = {tab:list_models}, 
]{
  width = \linewidth, 
  colspec = {X[2, l, m] X[1, l, m] l}, 
  rowhead = 1, 
  rowsep = 0ex,
  cells = {font=\scriptsize}
}
\toprule
\textbf{Hugging Face ID} & \textbf{Model Lineage} & \textbf{Type} \\
\midrule

Qwen/Qwen2.5-7B & Qwen2.5-7B & PT \\
Qwen/Qwen2.5-7B-Instruct & Qwen2.5-7B & IT \\
Qwen/Qwen2.5-Math-7B & Qwen2.5-7B & FT \\
Qwen/Qwen2.5-Coder-7B-Instruct & Qwen2.5-7B & FT \\
WangCa/Qwen2.5-7B-Medicine & Qwen2.5-7B & FT \\
huihui-ai/Qwen2.5-7B-Instruct-abliterated-v2 & Qwen2.5-7B & FT \\
Locutusque/StockQwen-2.5-7B & Qwen2.5-7B & MM \\
bunnycore/QevaCoT-7B-Stock & Qwen2.5-7B & MM \\
fangcaotank/task-10-Qwen-Qwen2.5-7B-Instruct & Qwen2.5-7B & PEFT \\
SeeFlock/task-12-Qwen-Qwen2.5-7B-Instruct & Qwen2.5-7B & PEFT \\
Qwen/Qwen2.5-7B-Instruct-GPTQ-Int4 & Qwen2.5-7B & QZ \\
Qwen/Qwen2.5-7B-Instruct-GPTQ-Int8 & Qwen2.5-7B & QZ \\
Lansechen/Qwen2.5-7B-Open-R1-Distill & Qwen2.5-7B & DT \\
\midrule

Qwen/Qwen2.5-14B & Qwen2.5-14B & PT \\
Qwen/Qwen2.5-14B-Instruct & Qwen2.5-14B & IT \\
Qwen/Qwen2.5-Coder-14B & Qwen2.5-14B & FT \\
oxyapi/oxy-1-small & Qwen2.5-14B & FT \\
v000000/Qwen2.5-14B-Gutenberg-Instruct-Slerpeno & Qwen2.5-14B & MM \\
ToastyPigeon/qwen-story-test-qlora & Qwen2.5-14B & PEFT \\
Qwen/Qwen2.5-14B-Instruct-GPTQ-Int4 & Qwen2.5-14B & QZ \\
deepseek-ai/DeepSeek-R1-Distill-Qwen-14B & Qwen2.5-14B & DT \\
\midrule

meta-llama/Llama-3.1-8B & Llama-3.1-8B & PT \\
meta-llama/Llama-3.1-8B-Instruct & Llama-3.1-8B & IT \\
ValiantLabs/Llama3.1-8B-Fireplace2 & Llama-3.1-8B & FT \\
RedHatAI/Llama-3.1-8B-tldr & Llama-3.1-8B & FT \\
proxectonos/Llama-3.1-Carballo & Llama-3.1-8B & FT \\
mlabonne/Meta-Llama-3.1-8B-Instruct-abliterated & Llama-3.1-8B & FT \\
gaverfraxz/Meta-Llama-3.1-8B-Instruct-HalfAbliterated-TIES & Llama-3.1-8B & MM \\
Xiaojian9992024/Llama3.1-8B-ExtraMix & Llama-3.1-8B & MM \\
LlamaFactoryAI/Llama-3.1-8B-Instruct-cv-job-description-matching & Llama-3.1-8B & PEFT \\
chchen/Llama-3.1-8B-Instruct-PsyCourse-fold7 & Llama-3.1-8B & PEFT \\
iqbalamo93/Meta-Llama-3.1-8B-Instruct-GPTQ-Q\_8 & Llama-3.1-8B & QZ \\
DaraV/LLaMA-3.1-8B-Instruct-INT4-GPTQ & Llama-3.1-8B & QZ \\
asas-ai/Llama-3.1-8B-Instruct-Open-R1-Distill & Llama-3.1-8B & DT \\
\midrule

mistralai/Mistral-7B-v0.3 & Mistral-7B-v0.3 & PT \\
mistralai/Mistral-7B-Instruct-v0.3 & Mistral-7B-v0.3 & IT \\
KurmaAI/AQUA-7B & Mistral-7B-v0.3 & FT \\
openfoodfacts/spellcheck-mistral-7b & Mistral-7B-v0.3 & FT \\
grimjim/Mistral-7B-Instruct-demi-merge-v0.3-7B & Mistral-7B-v0.3 & MM \\
chaymaemerhrioui/mistral-Brain\_Model\_ACC\_Trainer & Mistral-7B-v0.3 & PEFT \\
RedHatAI/Mistral-7B-Instruct-v0.3-GPTQ-4bit & Mistral-7B-v0.3 & QZ \\
eganwo/mistral7b-distilled-from-deepseek-r1-qwen32b & Mistral-7B-v0.3 & DT \\
\midrule

google/gemma-2-2b & Gemma-2-2b & PT \\
google/gemma-2-2b-it & Gemma-2-2b & IT \\
rinna/gemma-2-baku-2b & Gemma-2-2b & FT \\
anakin87/gemma-2-2b-neogenesis-ita & Gemma-2-2b & FT \\
vonjack/gemma2-2b-merged & Gemma-2-2b & MM \\
google-cloud-partnership/gemma-2-2b-it-lora-sql & Gemma-2-2b & PEFT \\
qilowoq/gemma-2-2B-it-4Bit-GPTQ & Gemma-2-2b & QZ \\
Syed-Hasan-8503/Gemma-2-2b-it-distilled & Gemma-2-2b & DT \\
\midrule

TinyLlama/TinyLlama-1.1B-Chat-v1.0 & TinyLlama-1.1B-v1.0 & IT \\
alexredna/TinyLlama-1.1B-Chat-v1.0-reasoning-v2 & TinyLlama-1.1B-v1.0 & FT \\
Edentns/DataVortexTL-1.1B-v0.1 & TinyLlama-1.1B-v1.0 & FT \\
appvoid/dot-v2.7 & TinyLlama-1.1B-v1.0 & MM \\
barissglc/tinyllama-tarot-v1 & TinyLlama-1.1B-v1.0 & PEFT \\
TheBloke/TinyLlama-1.1B-Chat-v1.0-GPTQ & TinyLlama-1.1B-v1.0 & QZ \\
anudaw/distilled-code-llama & TinyLlama-1.1B-v1.0 & DT \\
\midrule

meta-llama/Llama-2-7b-hf & Llama-2-7b & PT\\
meta-llama/Llama-2-7b-chat-hf & Llama-2-7b & IT \\
allenai/tulu-2-7b & Llama-2-7b & FT \\
QIAIUNCC/EYE-Llama\_qa & Llama-2-7b & FT \\
DevQuasar/coma-7B-v0.1 & Llama-2-7b & MM \\
Ammar-1/llama2-Better-Tune & Llama-2-7b & PEFT \\
TheBloke/Llama-2-7B-Chat-GPTQ & Llama-2-7b & QZ \\
cygu/llama-2-7b-logit-watermark-distill-kgw-k1-gamma0.25-delta2 & Llama-2-7b & DT \\
\bottomrule
\end{longtblr}

\end{document}